\documentclass[Journal,twocolumn]{IEEEtran}
\IEEEoverridecommandlockouts
\usepackage{color}
\usepackage{graphicx}
\usepackage{amsmath}
\usepackage{amssymb}
\usepackage{algorithm}
\usepackage{algorithmic}
\usepackage{amsmath}
\usepackage{multirow}
\usepackage{booktabs}
\usepackage{array}
\usepackage{amsthm}
\usepackage{stfloats}
\usepackage{caption}
\usepackage{subfigure}
\usepackage{bm}
\usepackage{booktabs}
\usepackage{setspace}
\usepackage{url}
\usepackage{arydshln}

\newcommand{\be}{\begin{equation}}
\newcommand{\ee}{\end{equation}}

\newcommand{\non}{\nonumber}

\renewcommand{\thesubsubsection}{\thesubsection.\arabic{subsubsection}}

\renewcommand{\baselinestretch}{1}
\allowdisplaybreaks[4]

\title{ 
DOA Estimation-Oriented Joint Array Partitioning and Beamforming Designs for ISAC Systems
\thanks{Part of this paper has been presented in IEEE Global Communications Conference, 2023 \cite{Rang Globecom2023}.}
\thanks{Manuscript received May 10, 2024; revised October 15, 2024; accepted December 5, 2024. The work of Rang Liu and A. Lee Swindlehurst was supported by the U.S. National Science Foundation under Grant CCF-2225575 and Grant CCF-2322191. The work of Ming Li and Qian Liu was supported by the National Natural Science Foundation of China under Grant 62371090 and Grant 62471086. The associate editor coordinating the review of this article and approving it for publication was Dr. Vasanthan Raghavan. \textit{(Corresponding author: Rang Liu.)}}
\thanks{R. Liu and A. L. Swindlehurst are with the Center for Pervasive Communications and Computing, University of California, Irvine, CA 92697, USA (e-mail: rangl2@uci.edu; swindle@uci.edu).}
\thanks{M. Li is with the School of Information and Communication Engineering, Dalian University of Technology, Dalian 116024, China (e-mail: mli@dlut.edu.cn).}
\thanks{Q. Liu is with the School of Computer Science and Technology, Dalian University of Technology, Dalian 116024, China (e-mail: qianliu@dlut.edu.cn).}}

\author{Rang Liu,~\IEEEmembership{Member,~IEEE,}
        Ming Li,~\IEEEmembership{Senior Member,~IEEE,}
        Qian Liu,~\IEEEmembership{Member,~IEEE,}\\
        and A. Lee Swindlehurst,~\IEEEmembership{Fellow,~IEEE}}

\begin{document}

\maketitle
\pagestyle{empty}
\thispagestyle{empty}

\begin{abstract}
Integrated sensing and communication has been identified as an enabling technology for forthcoming wireless networks. In an effort to achieve an improved performance trade-off between multiuser communications and radar sensing, this paper considers a dynamically-partitioned antenna array architecture for monostatic ISAC systems, in which each element of the array at the base station can function as either a transmit or receive antenna. To fully exploit the available spatial degrees of freedom for both communication and sensing functions, we jointly design the partitioning of the array between transmit and receive antennas together with the transmit beamforming in order to minimize the direction-of-arrival (DOA) estimation error, while satisfying constraints on the communication signal-to-interference-plus-noise ratio and the transmit power budget. An alternating algorithm based on Dinkelbach’s transform, the alternative direction method of multipliers, and majorization-minimization is developed to solve the resulting complicated optimization problem. To reduce the computational complexity, we also present a heuristic three-step strategy that optimizes the transmit beamforming after determining the antenna partitioning. Simulation results confirm the effectiveness of the proposed algorithms in significantly reducing the DOA estimation error.

\end{abstract}

\section{Introduction}

The evolution of communication networks not only strives for higher data rates but also integrates diverse capabilities to support various emerging services.
To achieve this goal, integrated sensing and communication (ISAC) has been deemed as one of the key enabling techniques for future communication networks \cite{Liu TCOM 2020}, \cite{Zhang ICST 2022}.
ISAC enables the sharing of spectral resources between radar sensing systems and wireless communication systems, and also facilitates their integration onto a single platform, thereby substantially improving spectral/hardware/energy efficiency \cite{Liu JSAC 2022}.
Moreover, ubiquitously deployed base stations (BS) and terminal devices have the ability to create a pervasive sensing network, which will enhance communication efficiency and bridge the gap between the tangible physical realm and the virtual digital domain. Attracted by these benefits, researchers from both academia and industry have recently been exploring potential ISAC implementations in forthcoming wireless networks \cite{An TWC 2023}-\cite{jLi TCOM 2024}.

ISAC systems commonly use multi-input multi-output (MIMO) architectures to enhance both communication and sensing capabilities, which require sophisticated beamforming designs to effectively exploit the available spatial degrees of freedom (DoFs) \cite{Liu CST 2022}, \cite{Liu SPM 2023}. Driven by the advantages of MIMO-ISAC systems, numerous advanced beamforming design algorithms have been proposed to meet specific sensing and communication requirements in ISAC systems \cite{Zhang-JSTSP-2021}-\cite{Xiao 2023}. 
For example, the radar sensing performance is typically evaluated by the transmit beampattern \cite{Liu TWC2018}, the signal-to-interference-plus-noise ratio (SINR) \cite{Liu-JSAC-2022}, the Cram\'{e}r-Rao bound (CRB) for parameter estimation \cite{Liu TSP 2022}, etc. Popular performance metrics for multiuser communications involve the SINR \cite{Liu-TSP-2020}, the sum-rate \cite{Rang TWC 2024}, the safety margin \cite{Liu-JSAC-2022}, etc.

In these prior works on MIMO-ISAC systems, the transmit/receive antenna array is predetermined and fixed for the corresponding beamforming designs. 
However, the use of a fixed transmit/receive array architecture limits the exploitation of the available spatial DoFs and may lead to performance deterioration in different environmental circumstances. 
Antenna selection strategies, wherein antenna elements are dynamically selected from a given array for transmission or reception, have been extensively studied as an alternative to fixed transmit/receive array structures in both MIMO communication and MIMO radar systems.

For massive MIMO communication systems, antenna selection is increasingly desirable to reduce the hardware/computational complexity and power consumption as antennas are becoming cheaper and smaller compared to the radio-frequency (RF) front-ends \cite{Sanayei-CM-2004}. 
Sparse array design is a typical antenna selection approach in which only a limited number of antenna elements are dynamically activated from a dense array for transmitting signals.
Advanced sparse antenna selection algorithms have been proposed to maximize the energy efficiency \cite{Li TCOM2014} or capacity \cite{Gao TSP 2018}.
Similarly, sparse antenna array designs are also advantageous in MIMO radar systems for expanding the array aperture or substantially reducing cost while maintaining performance. 
Numerous investigations have been dedicated to designing sparse antenna arrays in MIMO radar systems by optimizing the transmit beampattern  \cite{Wang-TSP-2021}, the SINR \cite{Wang TSP2017}, the mean squared error (MSE) of direction-of-arrival (DOA) estimation \cite{Chen-TVT-2022}, the CRB for DOA estimation \cite{Arash TWC 2023}, etc.

Given its well-established benefits in MIMO communication and MIMO radar systems, researchers have recently begun investigating the potential of employing antenna selection in MIMO-ISAC systems. The authors in \cite{Wang TAES2019} employed antenna selection as a means of embedding communication information into different transmit array configurations, while \cite{Huang Radar2023} similarly investigated embedding communication information into transmit beampatterns and jointly designed the transmit beamforming and antenna selection vector to achieve favorable performance. The authors in \cite{Valiulahi-WCL-2022} considered the power consumption per transmit antenna and attempted to minimize the CRB for DOA estimation under communication SINR requirements and a total power budget. 
In \cite{Xu-ICASSP-2022}, the weighted sum of the communication rate and the Fisher information matrix was maximized under a transmit power budget and a fixed number of activated transmit antennas. 
In \cite{Wang Radar2023}, the power illuminating the target was maximized for a given number of activated transmit antennas under the constraints of communication SINR, transmit power, and clutter power. 
Although the aforementioned studies verified the benefits of antenna selection in MIMO-ISAC systems, they all use the similar idea of adaptively selecting transmit antennas from a large array, while the receive antenna array is pre-determined. Hence, the potentially available spatial DoFs are not fully exploited, and this limits performance since the characteristics of the receive antenna array greatly influence the sensing capabilities of MIMO-ISAC systems.

Motivated by the above discussions, we advocate a more flexible \textit{array partitioning} architecture for MIMO-ISAC systems.
Unlike the antenna selection approaches considered in the existing literature, the proposed array partitioning architecture can employ each individual element as either a transmit or a receive antenna, thus more fully exploiting the available spatial DoFs of the shared antenna array at the dual-functional BS.
The use of an adaptively partitioned array leads to an intrinsic compromise that requires careful consideration for ISAC systems.   
From the perspective of the transmit antenna array, more transmit antennas provide greater beamforming gains and improved multiuser interference suppression, leading to higher communication SINR and better communication quality of service (QoS).
Moreover, increasing the number of transmit antennas enables a more precise beam direction toward the potential target and hence stronger illumination of the target, which in turn produces stronger target echoes at the receive antennas, contributing to more accurate estimation/detection results.
However, from the perspective of the receive antenna array, the availability of more receive antennas will enhance the DOA estimation performance, and the specific spatial distribution of the receive antennas also significantly influences the effective aperture and estimation accuracy.
Consequently, array partitioning, which involves the selection of both the number of transmit/receive antennas and their spatial distributions, is crucial for fully exploiting the available spatial DoFs, as it will simultaneously affect communication and sensing performance in different ways. Therefore, it is important to develop flexible array partitioning strategies and associated beamforming designs to achieve a flexible performance trade-off between multiuser communications and radar sensing for MIMO-ISAC systems.

This paper presents the first investigation of a joint array partitioning and transmit beamforming approach for monostatic MIMO-ISAC systems in which a multi-antenna BS simultaneously transmits dual-functional signals to serve multiple communication users and receives echo signals to estimate the target DOA. 
The main contributions of this paper are summarized as follows.
\begin{itemize} 
  \item First, we establish the array partitioning model for the considered monostatic MIMO-ISAC system. The proposed array partitioning architecture allows for more flexible spatial resource allocation by leveraging additional spatial DoFs that have not been previously exploited, thus significantly enhancing performance trade-offs. Based on the use of the SINR metric for multiuser communications and the DOA root mean squared error (RMSE) metric for target DOA estimation, we jointly optimize the array partitioning and transmit beamforming to minimize the RMSE of the DOA estimation, as well as satisfy the communication SINR requirements, the transmit power budget, the necessary number of transmit antennas, and the integer constraints for the array partitioning. This problem has not previously been considered in the literature.
  
  \item We propose an alternating algorithm to solve the resulting complicated non-convex optimization problem. By applying tailored transformations that exploit the unique structure of the joint array partitioning and beamforming design, we employ the algorithmic frameworks of Dinkelbach’s transform, alternative direction method of multipliers (ADMM), and majorization-minimization (MM) methods to decompose the original problem into several tractable sub-problems, which can be solved in sequence for each variable.
  
  \item In order to reduce the computational complexity, we further develop a three-step heuristic algorithm to efficiently solve the joint array partitioning and transmit beamforming design problem. This approach first determines the required number of transmit/receive antennas, selects the corresponding transmit antennas from the antenna array, and then optimizes the transmit beamforming based on the obtained array partitioning result.
  
  \item We provide simulation results to demonstrate that the proposed joint array partitioning and transmit beamforming design can achieve notable performance improvements in DOA estimation compared with conventional antenna array configurations. The simulation results illustrate the significance of the inherent compromise between the partitioning of transmit and receive antennas in monostatic MIMO-ISAC systems.
\end{itemize}

\emph{Notation}: Boldface lower-case and upper-case letters indicate column vectors and matrices, respectively. The symbols
$(\cdot)^*$, $(\cdot)^T$, $(\cdot)^H$, and $(\cdot)^{-1}$ denote the conjugate, transpose, transpose-conjugate, and inverse operations, respectively. The space of real and imaginary numbers is respectively represented by $\mathbb{C}$ and $\mathbb{R}$. The operators
$| a |$, $\| \mathbf{a} \|$, and $\| \mathbf{A} \|_F$ represent the magnitude of a scalar $a$, the norm of a vector $\mathbf{a}$, and the Frobenius norm of a matrix $\mathbf{A}$, respectively. Statistical expectation is denoted by
$\mathbb{E}\{\cdot\}$, $\text{Tr}\{\mathbf{A}\}$ takes the trace of the matrix $\mathbf{A}$, and $\text{diag}\{\mathbf{a}\}$ indicates the diagonal matrix whose diagonal elements are taken from $\mathbf{a}$. The real part of a complex number is given by $\Re\{\cdot\}$, and  
$\mathcal{S}_N^+$ represents the set of all $N$-dimensional complex positive semidefinite matrices.
Finally, we let $\mathbf{A}(i,j)$ denote the element in the $i$-th row and the $j$-th column of matrix $\mathbf{A}$, and $\mathbf{a}(i)$ denote the $i$-th element of vector $\mathbf{a}$.

\section{System Model and Problem Formulation}\label{sec:system model}

\begin{figure}[!t]
\centering
\includegraphics[width = 3 in]{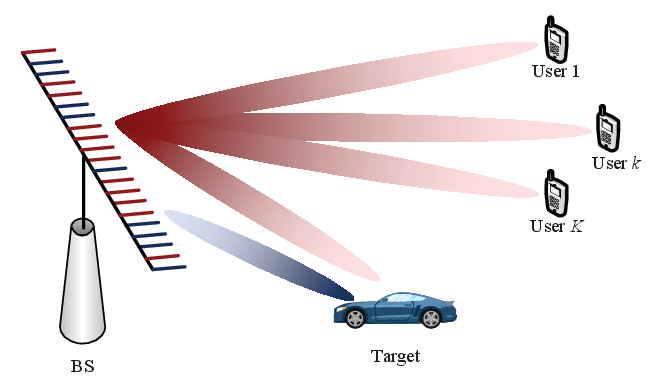}\vspace{0.1 cm}
\caption{A monostatic MIMO-ISAC system with a dynamically partitioned array (red: transmit antennas and beamforming, blue: receive antennas and beamforming).}
\label{fig:system_model}
\end{figure}

\subsection{System Model}
We consider a monostatic ISAC system as shown in Fig.~\ref{fig:system_model}, employing a BS equipped with $N$ antennas arranged as a uniform linear array (ULA) with half-wavelength spacing.
The monostatic ISAC BS simultaneously performs downlink multi-user communications and radar sensing.
Specifically, a subset of the elements within the antenna array is selected for the purpose of transmitting dual-functional signals that are intended to serve $K < N$ single-antenna communication users, while at the same time illuminating a single point-like target. The remaining antenna elements in the array are dedicated to receiving the echoes from the target, which are then used for DOA estimation\footnote{For clarity, we adopt a simplified system model with a ULA at the BS and single-antenna users. However, the proposed approach can be extended in a straightforward way to more complex scenarios involving planar arrays or multi-antenna users.}. 
In an effort to fully exploit the spatial DoFs and achieve a better performance trade-off between multiuser communications and target DOA estimation, the BS dynamically partitions the array into transmit and receive antennas and jointly optimizes the corresponding beamformers in accordance with different communication/sensing requirements and available resources.

We define the array partitioning vector as $\mathbf{a}\triangleq[a_1,a_2,\ldots,a_N]^T\in\{0,1\}^N$, where $a_n=1$ indicates that the $n$-th antenna is selected as a transmit antenna and $a_n=0$ denotes a receive antenna. The reconfiguration of the antennas is performed at a relatively low rate, typically no faster than the channel coherence time. RF switches can be used to selectively connect antenna elements to either a power amplifier for transmission or a low-noise amplifier for reception \cite{Rebeiz Microwave 2001}-\cite{Xia VTM 2018}, enabling the antenna elements to alternate between transmitting and receiving based on the current configuration vector $\mathbf{a}$.
The number of transmit antennas $N_\text{t}$ and receive antennas $N_\text{r}$ can be respectively calculated as
\begin{subequations}\begin{align}
N_\text{t} &= \mathbf{1}^T\mathbf{a},\\
N_\text{r} &= N-N_\text{t} = N-\mathbf{1}^T\mathbf{a}.\label{eq:Nr}
\end{align}\end{subequations}
Moreover, since the BS has to simultaneously transmit $K$ different data streams and receive the target echoes, the following constraint is imposed on $N_\text{t}$:
\be
K\leq N_\text{t}\leq N-1.
\ee

To simultaneously provide high-quality multiuser communications and target DOA estimation performance, the transmitted dual-functional waveform composed of precoded communication symbols and dedicated radar probing signals is expressed as \cite{Liu-TSP-2020}
\be
\mathbf{x} = \mathbf{A}\mathbf{W}_\text{c}\mathbf{s}_\text{c} + \mathbf{A}\mathbf{W}_\text{r}\mathbf{s}_\text{r} = \mathbf{A}\mathbf{W}\mathbf{s},
\ee
where $\mathbf{A} \triangleq \text{diag}\{\mathbf{a}\}$ denotes the array partitioning matrix, $\mathbf{s}_\text{c}\in\mathbb{C}^K$ denotes the transmitted communication symbols with $\mathbb{E}\{\mathbf{s}_\text{c}\mathbf{s}_\text{c}^H\}=\mathbf{I}_K$, $\mathbf{s}_\text{r}\in\mathbb{C}^N$ represents the radar probing waveform with $\mathbb{E}\{\mathbf{s}_\text{r}\mathbf{s}_\text{r}^H\}=\mathbf{I}_N$, and it is assumed that $\mathbf{s}_\text{c}$ and $\mathbf{s}_\text{r}$ have zero cross-correlation: $\mathbb{E}\{\mathbf{s}_\text{c}\mathbf{s}_\text{r}^H\}=\mathbf{0}$.
The matrices $\mathbf{W}_\text{c}\in\mathbb{C}^{N\times K}$ and $\mathbf{W}_\text{r}\in\mathbb{C}^{N\times N}$ are the beamformers for communications and radar sensing, respectively.
For notational simplicity, we define the integrated signal vector $\mathbf{s}  \triangleq [\mathbf{s}_\text{c}^T ~\mathbf{s}_\text{r}^T ]^T$  and the overall transmit beamforming matrix $\mathbf{W} \triangleq [\mathbf{W}_\text{c}~\mathbf{W}_\text{r}]\triangleq[\mathbf{w}_1, \mathbf{w}_2,\ldots, \mathbf{w}_{K+N}]$, where $\mathbf{w}_j$ represents the $j$-th column of matrix $\mathbf{W}$.
Although we formulate the transmit signal $\mathbf{x}$ as an $N$-dimension vector, only the elements corresponding to the transmit antennas have non-zero values, i.e., the $n$-th element of $\mathbf{x}$ is non-zero only if $a_n = 1$.

\subsection{MU-MISO Communications}
Based on the above model, the signal received at the $k$ -th user can be written as
\be
y_k = \mathbf{h}_{k}^T\mathbf{A}\mathbf{W}\mathbf{s} + n_k,
\ee
where $\mathbf{h}_{k}\in\mathbb{C}^N$ represents the channel between the $N$-antenna BS and the $k$-th user, and $n_k\sim\mathcal{CN}(0,\sigma_k^2)$ denotes additive white Gaussian noise (AWGN).
Given the availability of advanced channel estimation approaches for multi-user multiple-input single-output (MU-MISO) downlink communication systems, in this paper, we assume perfect channel state information (CSI) for $\mathbf{h}_k$. This assumption allows us to focus on analyzing the performance gains of the proposed array partitioning approach.
Thus, the SINR of the $k$-th communication user can be calculated as
\be
\text{SINR}_{k} = \frac{|\mathbf{h}_{k}^T\mathbf{A}\mathbf{w}_{k}|^2}
{\sum_{j\neq k}^{K+N}|\mathbf{h}_{k}^T\mathbf{A}\mathbf{w}_{j}|^2+\sigma_k^2},
\ee
which is a widely used metric for evaluating multiuser communication performance.

\subsection{Radar Sensing}

Meanwhile, the signals received at the BS, which consist of the target echos, the self-interference (SI) between the transmit and receive antennas, and the noise, can be expressed as
\be\label{eq:received echo signal}
\mathbf{y}_\text{r} =\! \alpha_\text{t}(\mathbf{I}_N-\mathbf{A})\mathbf{h}_\text{t}\mathbf{h}_\text{t}^T\mathbf{A}\mathbf{Ws} +
(\mathbf{I}_N-\mathbf{A})\mathbf{H}_\text{SI}\mathbf{A}\mathbf{Ws} + (\mathbf{I}_N-\mathbf{A})\mathbf{n}_\text{r},
\ee
where $\alpha_\text{t}\sim\mathcal{CN}(0,\sigma_\text{t}^2)$ is the radar cross section (RCS) of the target, $\mathbf{h}_\text{t}\in\mathbb{C}^N$ denotes the channel between the $N$-antenna BS and the target, $\mathbf{H}_\text{SI}\in\mathbb{C}^{N\times N}$ represents the SI channel between the transmit and receive antennas at the BS, and  $\mathbf{n}_\text{r}\sim\mathcal{CN}(\mathbf{0},\sigma_\text{r}^2\mathbf{I}_N)$ is AWGN.
It is obvious that echo signals can only be acquired by the antennas designated as receive antennas, i.e., the $n$-th element of $\mathbf{y}_\text{r}$ has a non-zero value only if $a_n = 0$.
In radar sensing applications, it is commonly assumed that the link between the BS and the target follows a line-of-sight (LoS) path, i.e., $\mathbf{h}_\text{t} = \beta_\text{t}\mathbf{h}(\theta_\text{t})$, where $\beta_\text{t}$ denotes the distance-dependent path-loss, $\theta_\text{t}$ represents the DOA of the target, and $\mathbf{h}(\theta)$ is the steering vector defined as 
\begin{equation}
\mathbf{h}(\theta) \triangleq [e^{-\jmath\frac{N-1}{2}\pi\sin\theta},e^{-\jmath\frac{N-3}{2}\pi\sin\theta}\ldots,e^{\jmath\frac{N-1}{2}\pi\sin\theta}]^T.  
\end{equation}

Here we consider a typical target tracking scenario in which the BS first estimates/predicts the target DOA and then steers its beam in the direction of the target to obtain a more accurate DOA estimate for improved tracking.
In this paper, we focus on the latter step, where the BS optimizes the transmit beamformer after obtaining a rough estimate of the target angle $\theta_\text{t}$.
We adopt the RMSE of the DOA estimation, denoted here as $\sigma_\theta$, to evaluate the radar sensing performance.
Theoretically, the DOA RMSE is determined by the 3dB beamwidth and the receive SINR, and is calculated as \cite{radar measurement}
\be\label{eq:sigma theta}
\sigma_\theta = \frac{\theta_\text{3dB}}{1.6\sqrt{2\text{SINR}_\text{r}}}.
\ee
According to the expression for the received echo signals in (\ref{eq:received echo signal}), the receive SINR can be calculated as 
\be\label{eq:SNRr}
\text{SINR}_\text{r} = \frac{\sigma_\text{t}^2\|(\mathbf{I}_N-\mathbf{A})\mathbf{h}_\text{t}\mathbf{h}_\text{t}^T\mathbf{A}\mathbf{W}\|_F^2}
{\sigma_\text{r}^2N_\text{r}+\|(\mathbf{I}_N-\mathbf{A})\mathbf{H}_\text{SI}\mathbf{A}\mathbf{W}\|_F^2}.
\ee

In addition, it is well-known that for a fully populated $N$-element antenna array with half-wavelength antenna spacing, the 3dB beamwidth is approximately $1.772/N$ if the beam is steered to boresight with uniform weights \cite{beamwidth}. 
The 3dB beamwidth will exceed $1.772/N$ in the general case where the beam is steered to an arbitrary angle $\theta_\text{t}$. Moreover, our considered array partitioning architecture further broadens the beamwidth, as not all $N$ elements are used for reception. Consequently, we define the 3dB beamwidth $\theta_\text{3dB}$ as 
\be\label{eq:3dB}
\theta_\text{3dB} = 1.772/N + 2\Delta,
\ee 
where $\Delta \geq 0$ represents the additional broadening caused by the off-boresight steering angle $\theta_\text{t}$ and the array partitioning vector $\mathbf{a}$. We should emphasize that the term $\Delta$ captures the nonlinear changes in the effective array aperture and 3dB beamwidth. Instead of an arbitrary or fixed additive error term, the scalar $\Delta$ is rigorously derived based on the array gain, providing an accurate and interpretable characterization of the variations in beamwidth under different steering angles and array partitioning strategies. Specifically, we denote the LoS channel corresponding to the half-power point as $\mathbf{h}\big(\theta_\text{t} + \frac{1}{2}\theta_\text{3dB}\big) = \mathbf{h}(\theta_0+\Delta)$, where $\theta_0 =\theta_\text{t} + 1.772/2/N$. 
According to the definition of 3dB beamwidth, the scalar $\Delta$ should satisfy
\be\label{eq:3db}
\frac{\big|\mathbf{h}^T(\theta_0+\Delta)(\mathbf{I}-\mathbf{A})\mathbf{h}^*(\theta_\text{t})\big|^2}{\big|\mathbf{h}^T(\theta_\text{t})(\mathbf{I}-\mathbf{A})\mathbf{h}^*(\theta_\text{t})\big|^2} = \frac{1}{2},
\ee
for which it is clear that $\Delta$ is a function of both $\theta_\text{t}$ and $\mathbf{a}$. 

In order to obtain a closed-form solution for $\Delta$, we employ a first-order Taylor expansion to approximate the LoS channel $\mathbf{h}(\theta_0+\Delta)$ as 
\begin{subequations}\begin{align}
\mathbf{h}(\theta_0+\Delta) &\approx \mathbf{h}(\theta_0) -\jmath\pi\Delta\cos\theta_0\mathbf{Q}\mathbf{h}(\theta_0) \\ \label{eq:hdt}
& = (\mathbf{I}-\jmath\pi\Delta\cos\theta_0\mathbf{Q})\mathbf{h}(\theta_0),
\end{align}\end{subequations}
where we use the first-order derivative of $\mathbf{h}(\theta)$ with respect to $\theta$ as ${\partial\mathbf{h}(\theta)}/{\partial\theta} = -\jmath\pi\cos\theta\mathbf{Q}\mathbf{h}(\theta)$ with $\mathbf{Q}\triangleq\text{diag}\{\mathbf{q}\}$ and $\mathbf{q} \triangleq [\frac{N-1}{2},\frac{N-3}{2},\ldots,-\frac{N-1}{2}]^T$. 
We note that the first-order Taylor expansion is employed at point $\theta_0$ since it is a lower bound for the half-power point, enabling a tighter approximation.
Using the approximation in (\ref{eq:hdt}), we have 
\begin{subequations}\begin{align}
&\big|\mathbf{h}^T(\theta_0+\Delta)(\mathbf{I}-\mathbf{A})\mathbf{h}^*(\theta_\text{t})\big|^2\nonumber
\\
& = \big|(\mathbf{1}-\jmath\pi\Delta\cos\theta_0\mathbf{q})^T\text{diag}\{\mathbf{h}(\theta_0)\}\text{diag}\{\mathbf{h}^*(\theta_\text{t})\}\mathbf{b}\big|^2\\
& = \Delta^2\pi^2\cos^2\!\theta_0|\mathbf{q}^T\!\text{diag}\{\mathbf{h}\}\mathbf{b}|^2 -\Delta\mathbf{b}^T\!\mathbf{Q}_0\mathbf{b} + |\mathbf{b}^T\!\mathbf{h}|^2,\! \label{eq:num 3db}
\end{align}\end{subequations}
where we define $\mathbf{b}\triangleq \mathbf{1}-\mathbf{a}$ and $\mathbf{h} \triangleq \text{diag}\{\mathbf{h}(\theta_0)\}\mathbf{h}^*(\theta_\text{t})$ for notational simplicity. The matrix $\mathbf{Q}_0 \triangleq\pi\cos\theta_0 \Im\{\mathbf{h}^*\mathbf{h}^T\mathbf{Q}+\mathbf{Qhh}^H\}$ is obviously a low-rank symmetric matrix whose trace equals zero. Thus, we can use an eigenvalue decomposition (EVD) to re-construct $\mathbf{Q}_0$ as $\mathbf{Q}_0 = \mathbf{Q}_2-\mathbf{Q}_1$, where $\mathbf{Q}_1$ and $\mathbf{Q}_2$ are positive definite matrices. In addition, using the structure of the steering vector $\mathbf{h}(\theta)$, the denominator of the left-hand side in (\ref{eq:3db}) is re-written as 
\be\big|\mathbf{h}^T(\theta_\text{t})(\mathbf{I}-\mathbf{A})\mathbf{h}^*(\theta_\text{t})\big|^2 = |\mathbf{b}^T\mathbf{1}|^2. \label{eq:deno 3db}
\ee
Substituting (\ref{eq:num 3db}) and (\ref{eq:deno 3db}) into (\ref{eq:3db}), we have the following equality with respect to $\Delta$:
\be\begin{aligned}
&\Delta^2\pi^2\cos^2\theta_0\big|\mathbf{b}^T\text{diag}\{\mathbf{h}^H\}\mathbf{q}\big|^2 - \Delta\mathbf{b}^T(\mathbf{Q}_2-\mathbf{Q}_1)\mathbf{b} \\
&\qquad + |\mathbf{b}^T\mathbf{h}|^2- \frac{1}{2}|\mathbf{b}^T\mathbf{1}|^2=0.
\end{aligned}\ee
We assume that the value of $\Delta$ is sufficiently small so that the quadratic term in $\Delta$ can be ignored, and thus our expression for $\Delta$ becomes 
\be\label{eq:Delta}
\Delta = \frac{2|\mathbf{b}^T\mathbf{h}|^2-|\mathbf{b}^T\mathbf{1}|^2}{2\mathbf{b}^T(\mathbf{Q}_2-\mathbf{Q}_1)\mathbf{b}}.
\ee

Substituting the result from (\ref{eq:SNRr}), (\ref{eq:3dB}) and (\ref{eq:Delta}) into (\ref{eq:sigma theta}), the DOA RMSE can be formulated as 
\be\label{eq:new sigma theta}
\sigma_\theta = \frac{0.7831+0.8839\Delta N}{N\sqrt{\text{SINR}_\text{r}}} = \frac{c_1+c_2\frac{2|\mathbf{b}^T\mathbf{h}|^2-|\mathbf{b}^T\mathbf{1}|^2}{\mathbf{b}^T(\mathbf{Q}_2-\mathbf{Q}_1)\mathbf{b}}}{\sqrt{\text{SINR}_\text{r}}},
\ee
where we define $c_1 = 0.7831/N$ and $c_2 = 0.4419$ for simplicity. 
It is evident that the DOA RMSE is dependent not only on the receive SINR but also on the array partitioning vector, which necessitates a joint design of the antenna partitioning vector $\mathbf{a}$ and the transmit beamforming matrix $\mathbf{W}$. 

\subsection{Problem Formulation}

In this paper, we propose to jointly design the array partitioning vector $\mathbf{a}$ and the transmit beamforming matrix $\mathbf{W}$ to minimize the DOA RMSE while satisfying the communication SINR requirements, the transmit power budget, and the constraints on the array partitioning vector.
The complete optimization problem is formulated as
\begin{subequations}\label{eq:original problem}\begin{align}
&\underset{\mathbf{W}, \mathbf{a}}\min~~\sigma_\theta \label{eq:original problem obj}\\
&~\text{s.t.}~~\text{SINR}_{k}\geq \Gamma_k,~\forall k, \label{eq:op c1}\\
&\qquad~\|\mathbf{AW}\|_F^2 \leq P,\\
&\qquad~K\leq\mathbf{1}^T\mathbf{a}\leq N-1,\\
&\qquad~a_n\in\{0,1\},~\forall n,\label{eq:op c4}
\end{align}\end{subequations}
where $\Gamma_k$ denotes the communication SINR threshold and $P$ is the transmit power budget. 
The non-convex fractional objective function (\ref{eq:original problem obj}), the binary integer constraint (\ref{eq:original problem}e), and the coupled variables present significant difficulties for solving this problem. 
To address these challenges, in the next section we propose to decompose this problem into several solvable sub-problems and tackle them iteratively.

\section{Joint Array Partitioning and Transmit Beamforming Design}\label{sec:Algorithm 1}

In this section, we develop an alternating algorithm to solve the joint array partitioning and transmit beamforming design problem in (\ref{eq:original problem}).
We first simplify the objective function and utilize Dinkelbach's transform to obtain a favorable objective function, and then convert the binary integer constraint into a quadratic penalty term with a box constraint.
Next, we introduce an auxiliary variable to implement ADMM and alternatively solve each sub-problem by finding a convex surrogate function using the MM method. 
The details of the proposed algorithm are described as follows.

\subsection{Objective Function Simplification}

It is clear that the complicated objective function greatly hinders the algorithm development.
Thus, before solving this problem, we first simplify the SINR term in the objective function (\ref{eq:original problem}a) and convert it into a more favorable one using the following transformations. 

We first rearrange the numerator of (\ref{eq:SNRr}) into a more compact form by leveraging the unique features of the array partitioning vector $\mathbf{a}$.
Recalling that $\mathbf{A}\triangleq\text{diag}\{\mathbf{a}\}$ and
\be
\mathbf{H}_\text{t} \triangleq (\mathbf{I}_N-\mathbf{A})\mathbf{h}_\text{t}\mathbf{h}_\text{t}^T\mathbf{A} = \text{diag}\{\mathbf{h}_\text{t}\}\mathbf{ba}^T\text{diag}\{\mathbf{h}_\text{t}\}, 
\ee
the term $\mathbf{H}^H_\text{t}\mathbf{H}_\text{t}$ can be converted to
\begin{subequations}\label{eq:HaHa}\begin{align}
\mathbf{H}^H_\text{t}\mathbf{H}_\text{t} &\overset{(\text{a})}{=} \text{diag}\{\mathbf{h}^*_\text{t}\}\mathbf{ab}^T\text{diag}\{\mathbf{h}^*_\text{t}\}
\text{diag}\{\mathbf{h}_\text{t}\}\mathbf{ba}^T\text{diag}\{\mathbf{h}_\text{t}\} \\
& \overset{(\text{b})}{=}  \beta^2_\text{t}\text{diag}\{\mathbf{h}^*_\text{t}\}\mathbf{ab}^T\mathbf{ba}^T\text{diag}\{\mathbf{h}_\text{t}\}\\
&\overset{(\text{c})}{=}  \beta^2_\text{t}N_\text{r}\text{diag}\{\mathbf{h}^*_\text{t}\}\mathbf{a}\mathbf{a}^T\text{diag}\{\mathbf{h}_\text{t}\},
\end{align}\end{subequations}
where (a) holds since each element of $\mathbf{a}$ and $\mathbf{b}$ is in $\{0,1\}$, (b) holds since the elements of $\mathbf{h}_\text{t}$ are on the circle and thus
$\text{diag}\{\mathbf{h}^*_\text{t}\}
\text{diag}\{\mathbf{h}_\text{t}\} = \beta^2_\text{t}\mathbf{I}_N$, and (c) holds since $\mathbf{b}^T\mathbf{b} = \mathbf{1}^T\mathbf{b} = N_\text{r}$.
Using the result in (\ref{eq:HaHa}c), the numerator of (\ref{eq:SNRr}) can be rewritten as
\begin{subequations}\begin{align}
&\big\|(\mathbf{I}_N-\mathbf{A})\mathbf{h}_\text{t}\mathbf{h}_\text{t}^T\mathbf{A}\mathbf{W}\big\|^2_F
		\\
&=\text{Tr}\big\{\mathbf{H}^H_\text{t}\mathbf{H}_\text{t}\mathbf{WW}^H\big\}\\
& = \beta^2_\text{t}N_\text{r}\text{Tr}\big\{\text{diag}\{\mathbf{h}^*_\text{t}\}\mathbf{a}\mathbf{a}^T
		\text{diag}\{\mathbf{h}_\text{t}\}\mathbf{WW}^H\big\}\\
&=\beta^2_\text{t}N_\text{r}\mathbf{a}^T
		\text{diag}\{\mathbf{h}_\text{t}\}\mathbf{WW}^H\text{diag}\{\mathbf{h}^*_\text{t}\}\mathbf{a}.\label{eq:transf num}
\end{align}\end{subequations}

For the denominator of (\ref{eq:SNRr}), we will use the following upper-bound approximation: 
\begin{subequations}\begin{align}
\|(\mathbf{I}_N-\mathbf{A})\mathbf{H}_\text{SI}\mathbf{A}\mathbf{W}\|_F^2 
&\leq \|\mathbf{I}_N-\mathbf{A}\|_F^2\|\mathbf{H}_\text{SI}\mathbf{A}\mathbf{W}\|_F^2 \\
&= N_\text{r}\|\mathbf{H}_\text{SI}\mathbf{A}\mathbf{W}\|_F^2 .\label{eq:transf deno}
\end{align}\end{subequations}
This approximation allows us to bring the denominator into a similar form as the numerator, facilitating an easier solution. 
In particular, substituting the results in (\ref{eq:transf num}) and (\ref{eq:transf deno}) into (\ref{eq:SNRr}), a lower bound for $\text{SINR}_\text{r}$ is given by  
\begin{equation}\label{eq:SINR lb}
\text{SINR}_\text{r} \geq \frac{\sigma_\text{t}^2\beta_\text{t}^2\mathbf{a}^T\text{diag}\{\mathbf{h}_\text{t}\}\mathbf{WW}^H\text{diag}\{\mathbf{h}^*_\text{t}\}\mathbf{a}}
{\sigma_\text{r}^2+\|\mathbf{H}_\text{SI}\mathbf{A}\mathbf{W}\|_F^2}.
\end{equation}
This concise fractional term together with the fractional expression for $\Delta$ restricts the applicability of numerous established algorithms designed for polynomial functions. To address this challenge, we use Dinkelbach's transform to convert the objective function into a more manageable form, as described below.

\subsection{Dinkelbach’s Transform} 
In order to handle the fractional terms associated with $\Delta$ and the SINR, we introduce two auxiliary variables $t_1\in\mathbb{R}$ and $t_2\in\mathbb{R}$, and then transform the optimization problem into  
\begin{subequations}\label{eq:obj3}\begin{align}
&\underset{\mathbf{W}, \mathbf{a}, t_1, t_2}\min~~\frac{c_1+c_2t_1}{\sqrt{t_2}}\\
&\quad~\text{s.t.}\quad~\frac{2|\mathbf{b}^T\mathbf{h}|^2-|\mathbf{b}^T\mathbf{1}|^2}{\mathbf{b}^T(\mathbf{Q}_2-\mathbf{Q}_1)\mathbf{b}} \leq t_1,\\
&\qquad\qquad \frac{\sigma_\text{t}^2\beta_\text{t}^2\mathbf{a}^T\text{diag}\{\mathbf{h}_\text{t}\}\mathbf{WW}^H\text{diag}\{\mathbf{h}^*_\text{t}\}\mathbf{a}}
{\sigma_\text{r}^2+\|\mathbf{H}_\text{SI}\mathbf{A}\mathbf{W}\|_F^2} \geq t_2,\\
&\qquad\qquad (\ref{eq:original problem}\text{b}) \sim (\ref{eq:original problem}\text{e}).\non
\end{align}\end{subequations}
The auxiliary variables $t_1$ and $t_2$ are alternately updated along with the other variables. According to Dinkelbach's transform \cite{Dinkelbach}, their optimal solutions in each iteration are given by
\begin{subequations}\label{eq:update t12}
\begin{align}
t_1^\star &= \frac{2|\mathbf{b}^T\mathbf{h}|^2-|\mathbf{b}^T\mathbf{1}|^2}{\mathbf{b}^T(\mathbf{Q}_2-\mathbf{Q}_1)\mathbf{b}},\\
t_2^\star &= \frac{\sigma_\text{t}^2\beta_\text{t}^2\mathbf{a}^T\text{diag}\{\mathbf{h}_\text{t}\}\mathbf{WW}^H\text{diag}\{\mathbf{h}^*_\text{t}\}\mathbf{a}}
{\sigma_\text{r}^2+\|\mathbf{H}_\text{SI}\mathbf{A}\mathbf{W}\|_F^2} .
\end{align}\end{subequations}
Note that the variable $t_2$ will be positive, while the sign of $t_1$ is indeterminate.

\subsection{Binary Integer Constraints} 
A direct implementation of the binary integer constraint in (\ref{eq:original problem}e) would involve a computationally prohibitive search over $2^N$ discrete points, so we propose instead to replace it with a more tractable constraint.
It is clear that any vector that satisfies the binary constraint $a_n\in\{0,1\},~\forall n$, is also an optimal solution to the following problem:
\begin{subequations}\label{eq:problem convert a}\begin{align}
&\underset{\mathbf{a}}\min~~\mathbf{a}^T(\mathbf{1}-\mathbf{a})\\
&~\text{s.t.}~~0\leq a_n \leq 1,~\forall n.
\end{align}\end{subequations}
By employing this finding, the constraint in (\ref{eq:original problem}e) can be eliminated by incorporating (\ref{eq:problem convert a}a) into the objective function (\ref{eq:obj3}a) as a penalty and introducing the box constraint (\ref{eq:problem convert a}b), as follows:
\begin{subequations}\label{eq:problem after relax a}\begin{align}
&\underset{\mathbf{W},\mathbf{a}, t_1,t_2}\min~~\frac{c_1+c_2t_1}{\sqrt{t_2}}+\rho_1\mathbf{a}^T(\mathbf{1}-\mathbf{a})\label{eq:problem after relax a obj} \\
&\quad~\text{s.t.}\quad~~(\ref{eq:original problem}\text{b})\sim(\ref{eq:original problem}\text{d}),~(\ref{eq:obj3}\text{b}),~(\ref{eq:obj3}\text{c}),\nonumber\\
&\qquad\qquad~ 0\leq a_n \leq 1,~\forall n,
\end{align}\end{subequations}
where $\rho_1>0$ is a penalty parameter to control the trade-off between the objective value and the degree to which the binary integer constraints are satisfied. The solution to (\ref{eq:problem after relax a}) has been demonstrated to satisfy the binary constraint as $\rho_1$ grows large enough \cite{Dinh-2010}. However, the value of $\rho_1$ is typically set to be of the same order of magnitude as the initial objective function in (\ref{eq:obj3}). This is done to prevent stagnation of the optimization process when $\rho_1$ is too large, and to avoid large deviations from the binary constraint that occur if $\rho_1$ is too small.

\subsection{ADMM-based Transformation}

Now we observe that the complicated constraints (\ref{eq:obj3}\text{b}) and (\ref{eq:obj3}\text{c}) are the main difficulties in solving the problem. 
To address this issue, we introduce an auxiliary variable $\mathbf{b}$ (as already done before for notational simplicity), and employ the ADMM framework to decouple these two constraints. 
In particular, an auxiliary variable $\mathbf{b} \in\mathbb{R}^N$ is introduced to transform problem (\ref{eq:problem after relax a}) as
\begin{subequations}\label{eq:problem after alpha}\begin{align}
&\underset{\mathbf{W},\mathbf{a},\mathbf{b}, t_1,t_2}\min~~\frac{c_1+c_2t_1}{\sqrt{t_2}}+\rho_1\mathbf{a}^T(\mathbf{1}-\mathbf{a})\\
&\quad~~\text{s.t.}\quad~~(\ref{eq:original problem}\text{b})\sim(\ref{eq:original problem}\text{d}),~(\ref{eq:obj3}\text{b}),~(\ref{eq:obj3}\text{c}),\\
&\qquad\qquad~~ 0\leq a_n, b_n \leq 1,~\forall n,\\
&\qquad\qquad~~1\leq\mathbf{1}^T\mathbf{b}\leq N-K,\\
&\qquad\qquad~~\mathbf{b} = \mathbf{1}-\mathbf{a},
\end{align}\end{subequations}
where constraint (\ref{eq:obj3}\text{b}) is with respect to the variable $\mathbf{b}$ and constraint (\ref{eq:obj3}\text{c}) is related to the variables $\mathbf{a}$ and $\mathbf{W}$. 
To use the ADMM framework  \cite{ADMM}, we define an indicator function $\mathbb{I}_\mathcal{C}(\mathbf{W}, \mathbf{a},\mathbf{b}, t_1, t_2)$ that represents the feasible region defined by constraints (\ref{eq:problem after alpha}b)-(\ref{eq:problem after alpha}d) as 
\begin{equation}\label{eq:indicator function}
\mathbb{I}_\mathcal{C}(\mathbf{W}, \mathbf{a},\mathbf{b}, t_1, t_2) = \left\{
\begin{array}{lr}0,\hspace{0.6 cm} (\ref{eq:problem after alpha}\text{b})-(\ref{eq:problem after alpha}\text{d})\text{ are satisfied},\\
+\infty,\hspace{0.2 cm}\text{otherwise}.
\end{array}
\right.
\end{equation}
Using $\mathbb{I}_\mathcal{C}(\mathbf{W}, \mathbf{a},\mathbf{b}, t_1, t_2)$ to impose constraints (\ref{eq:problem after alpha}b)-(\ref{eq:problem after alpha}d) on the objective function, problem (\ref{eq:problem after alpha}) can be rewritten as
\begin{subequations}\label{eq:problem after alpha with indicator}\begin{align}
&\underset{\mathbf{W},\mathbf{a},\mathbf{b}, t_1, t_2}\min~~\frac{c_1+c_2t_1}{\sqrt{t_2}}+\rho_1\mathbf{a}^T(\mathbf{1}-\mathbf{a}) + \mathbb{I}_\mathcal{C}(\mathbf{W}, \mathbf{a}, \mathbf{b}, t_1, t_2)\\
&\quad~~ \text{s.t.}\qquad\mathbf{b} = \mathbf{1}-\mathbf{a}.
\end{align}\end{subequations}

The optimal solution to (\ref{eq:problem after alpha with indicator}) is obtained by minimizing its augmented Lagrangian (AL) function:
\be\begin{aligned}\label{eq:original AL function}
&\mathcal{L}(\mathbf{W},\mathbf{a},\mathbf{b},t_1,t_2,\bm{\mu})\triangleq\frac{c_1+c_2t_1}{\sqrt{t_2}}+\rho_1\mathbf{a}^T(\mathbf{1}-\mathbf{a}) \\
&\qquad+ \mathbb{I}_\mathcal{C}(\mathbf{W}, \mathbf{a}, \mathbf{b}, t_1, t_2) + \rho_2\big\|\mathbf{a}+\mathbf{b}-\mathbf{1}+{\bm{\mu}}/{\rho_2} \big\|^2 ,
\end{aligned}\ee
where $\rho_2>0$ is a penalty parameter and $\bm{\mu}\in\mathbb{R}^N$ is the dual variable.
Then, the block coordinate descent (BCD) method is employed to minimize the multivariate AL function by alternately updating each variable.
However, the non-concave terms in $\mathcal{L}(\mathbf{W},\mathbf{a},\mathbf{b}, t_1, t_2,\bm{\mu})$ lead to difficulties in solving the sub-problems.
To address this issue, in the following subsection we employ the MM method to construct a sequence of solvable sub-problems to be optimized until convergence \cite{Sun TSP 17}. 
The details for the updates of each variable are presented in the next subsection.

\subsection{MM-based Update}

\subsubsection{Update $\mathbf{W}$}
Given the solutions $\mathbf{W}^{(m)}$, $\mathbf{a}^{(m)}$, $\mathbf{b}^{(m)}$, $t_1^{(m)}$, $t_2^{(m)}$, and $\bm{\mu}^{(m)}$ obtained in the $m$-th iteration, the transmit beamformer $\mathbf{W}$ is updated by minimizing the AL function $\mathcal{L}(\mathbf{W},\mathbf{a}^{(m)},\mathbf{b}^{(m)}, t_1^{(m)}, t_2^{(m)},\bm{\mu}^{(m)})$, which is equivalent to 
\begin{subequations}\label{eq:problem w}\begin{align}
&\text{Find}~~ \mathbf{W}\\
&~\text{s.t.}~~\frac{\sigma_\text{t}^2\beta_\text{t}^2\mathbf{a}^T\text{diag}\{\mathbf{h}_\text{t}\}\mathbf{WW}^H\text{diag}\{\mathbf{h}^*_\text{t}\}\mathbf{a}}
{\sigma_\text{r}^2+\|\mathbf{H}_\text{SI}\mathbf{A}\mathbf{W}\|_F^2} \geq t_2,\\
&\qquad~\frac{|\mathbf{h}_{k}^T\mathbf{A}\mathbf{w}_{k}|^2}
{\sum_{j\neq k}^{K+N}|\mathbf{h}_{k}^T\mathbf{A}\mathbf{w}_{j}|^2+\sigma_k^2} \geq \Gamma_k,~\forall k,\\
&\qquad~\|\mathbf{A}\mathbf{W}\|_F^2 \leq P.
\end{align}\end{subequations}
Problem (\ref{eq:problem w}) is a feasibility-check without an explicit objective function. 
In order to accelerate the convergence and leave more flexibility for the following iterations, we propose to maximize the left-hand side of the constraint (\ref{eq:problem w}\text{b}), i.e.,
\begin{subequations}\label{eq:problem w2}\begin{align}
&\underset{\mathbf{W}}\max~~\mathbf{a}^T\text{diag}\{\mathbf{h}_\text{t}\}\mathbf{WW}^H\text{diag}\{\mathbf{h}^*_\text{t}\}\mathbf{a}-\frac{t_2}{\sigma_\text{t}^2\beta_\text{t}^2}\|\mathbf{H}_\text{SI}\mathbf{A}\mathbf{W}\|_F^2 \label{eq:problem w2a}\\
&~\text{s.t.}~~(\ref{eq:problem w}\text{c}),~(\ref{eq:problem w}\text{d}).\nonumber
\end{align}\end{subequations}

We observe that the objective function is non-concave due to the convex quadratic term $\mathbf{a}^T\text{diag}\{\mathbf{h}_\text{t}\}\mathbf{WW}^H\text{diag}\{\mathbf{h}^*_\text{t}\}\mathbf{a}$. 
Since a conditionally concave surrogate function is favorable in solving for each variable, we propose to seek a linear surrogate function for this convex term.
In particular, by employing a first-order Taylor expansion, a surrogate function for $\mathbf{a}^T\text{diag}\{\mathbf{h}_\text{t}\}\mathbf{WW}^H\text{diag}\{\mathbf{h}^*_\text{t}\}\mathbf{a}$ at the current point $\{\mathbf{W}^{(m)},\mathbf{a}^{(m)}\}$ is given by 
\be\begin{aligned}\label{eq:surr1}
&\mathbf{a}^T\text{diag}\{\mathbf{h}_\text{t}\}\mathbf{WW}^H\text{diag}\{\mathbf{h}^*_\text{t}\}\mathbf{a}
\\
&\qquad\geq
2\Re\big\{\mathbf{a}^T\text{diag}\{\mathbf{h}_\text{t}\}\mathbf{W}\mathbf{d}^{(m)} \big\} - \|\mathbf{d}^{(m)}\|^2,
\end{aligned}\ee
where we define $\mathbf{d}^{(m)} \triangleq (\mathbf{W}^{(m)})^H\text{diag}\{\mathbf{h}^*_\text{t}\}\mathbf{a}^{(m)}$. We note that (\ref{eq:surr1}) provides a linear surrogate function with respect to $\mathbf{W}$ and $\mathbf{a}$, enabling various convex optimization methods. 
Similarly, a linear surrogate function for the quadratic convex term $|\mathbf{h}_k^T\mathbf{Aw}_k|^2$ in constraint (\ref{eq:problem w}c) can be constructed as 
\begin{equation}\label{eq:hwk surr}
|\mathbf{h}_k^T\mathbf{Aw}_k|^2 \geq 2\Re\{\mathbf{h}_k^T\mathbf{Aw}_ke_k^{(m)}\} - |e_k^{(m)}|^2,
\end{equation}
where we define $e_k^{(m)}\triangleq (\mathbf{w}_k^{(m)})^H\text{diag}\{\mathbf{a}^{(m)}\}\mathbf{h}_k^*$.

Using the results in (\ref{eq:surr1}) and (\ref{eq:hwk surr}) and ignoring irrelevant terms, the sub-problem for finding $\mathbf{W}$ becomes
\begin{subequations}\label{eq:solve for W}\begin{align}
&\underset{\mathbf{W}}\max~~2\Re\{\mathbf{a}^T\text{diag}\{\mathbf{h}_\text{t}\}\mathbf{W}\mathbf{d}^{(m)}\}-\frac{t_2}{\sigma_\text{t}^2\beta_\text{t}^2}\|\mathbf{H}_\text{SI}\mathbf{AW}\|_F^2\\
&~\text{s.t.}~~2\Re\{\mathbf{h}_k^T\mathbf{Aw}_k{e}_k^{(m)}\} - |{e}_k^{(m)}|^2 \nonumber\\
&\qquad\qquad -\Gamma_k\sum_{j\neq k}^{K+N}|\mathbf{h}_{k}^T\mathbf{A}\mathbf{w}_{j}|^2-\Gamma_k\sigma_k^2 \geq 0,~\forall k,\\
&\qquad\quad \|\mathbf{AW}\|_F^2 \leq P,
\end{align}\end{subequations}
which is a convex quadratically constrained quadratic program (QCQP) problem that can be readily solved by various off-the-shelf algorithms or toolboxes, e.g., CVX \cite{cvx}.

\subsubsection{Update $\mathbf{a}$}
With the obtained solutions $\mathbf{W}^{(m+1)}$, $\mathbf{a}^{(m)}$, $\mathbf{b}^{(m)}$, $t_1^{(m)}$, $t_2^{(m)}$, and $\bm{\mu}^{(m)}$, the sub-problem of minimizing the AL function $\mathcal{L}(\mathbf{W}^{(m+1)},\mathbf{a},\mathbf{b}^{(m)}, t_1^{(m)}, t_2^{(m)},\bm{\mu}^{(m)})$ to update $\mathbf{a}$ can be expressed as 
\begin{subequations}\label{eq:subproblem a}\begin{align}
&\underset{\mathbf{a}}\min~~\rho_1\mathbf{a}^T(\mathbf{1}-\mathbf{a})+\rho_2 \|\mathbf{a}+\mathbf{b}-\mathbf{1}+\bm{\mu}/\rho_2 \|^2\\
&~~\text{s.t.}~~\frac{\sigma_\text{t}^2\beta_\text{t}^2\mathbf{a}^T\text{diag}\{\mathbf{h}_\text{t}\}\mathbf{WW}^H\text{diag}\{\mathbf{h}^*_\text{t}\}\mathbf{a}}
{\sigma_\text{r}^2+\|\mathbf{H}_\text{SI}\mathbf{A}\mathbf{W}\|_F^2} \geq t_2,\\
&\qquad~~\frac{|\mathbf{h}_{k}^T\mathbf{A}\mathbf{w}_{k}|^2}
{\sum_{j\neq k}^{K+N}|\mathbf{h}_{k}^T\mathbf{A}\mathbf{w}_{j}|^2+\sigma_k^2} \geq \Gamma_k,~\forall k,\\
&\qquad~~\|\mathbf{AW}\|_F^2 \leq P, \\
&\qquad~~K\leq\mathbf{1}^T\mathbf{a}\leq N-1,\\
&\qquad~~0\leq a_n \leq 1,~\forall n.
\end{align}\end{subequations}
A linear surrogate function for the convex term $\mathbf{a}^T\mathbf{a}$ in the objective is favorable, and again we construct such a function using a
first-order Taylor expansion as follows:
\be\label{eq:surr2}
\mathbf{a}^T\mathbf{a} \geq 2\mathbf{a}^T\mathbf{a}^{(m)}-\|\mathbf{a}^{(m)}\|^2.\ee

Meanwhile, the linear surrogate functions derived in (\ref{eq:surr1}) and (\ref{eq:hwk surr}) can be used to convexify the constraints in (\ref{eq:subproblem a}b) and (\ref{eq:subproblem a}c). 
Based on the results in  (\ref{eq:surr1}), (\ref{eq:hwk surr}), and (\ref{eq:surr2}), sub-problem (\ref{eq:subproblem a}) that solves for $\mathbf{a}$ is transformed into a convex QCQP problem as follows:
\begin{subequations}\label{eq:solve for a}\begin{align}
&\underset{\mathbf{a}}\min~~\rho_1\mathbf{a}^T(\mathbf{1}-2\mathbf{a}^{(m)})+\rho_2 \|\mathbf{a}+\mathbf{b}-\mathbf{1}+\bm{\mu}/\rho_2 \|^2\\
&~\text{s.t.}~~2\sigma_\text{t}^2\beta_\text{t}^2\Re\{\mathbf{a}^T\text{diag}\{\mathbf{h}_\text{t}\}\mathbf{W}\mathbf{d}^{(m)}\}-\sigma_\text{t}^2\beta_\text{t}^2\|\mathbf{d}^{(m)}\|^2 \nonumber\\
&\qquad\qquad\quad - t_2\sigma_\text{r}^2-t_2\|\mathbf{H}_\text{SI}\mathbf{AW}\|_F^2 \geq 0,\\
&\qquad~ 2\Re\{\mathbf{h}_k^T\mathbf{Aw}_k{e}_k^{(m)}\} - |{e}_k^{(m)}|^2 \nonumber\\
&\qquad\qquad\quad -\Gamma_k\sum_{j\neq k}^{K+N}|\mathbf{h}_{k}^T\mathbf{A}\mathbf{w}_{j}|^2-\Gamma_k\sigma_k^2 \geq 0,~\forall k,\\
&\qquad~\|\mathbf{AW}\|_F^2 \leq P, \\
&\qquad~K\leq\mathbf{1}^T\mathbf{a}\leq N-1,\\
&\qquad~0\leq a_n \leq 1,~\forall n,
\end{align}\end{subequations}
which can be readily solved by various algorithms. 

\subsubsection{Update $\mathbf{b}$} 

Given the solutions $\mathbf{W}^{(m+1)}$, $\mathbf{a}^{(m+1)}$, $\mathbf{b}^{(m)}$, $t_1^{(m)}$, $t_2^{(m)}$, and $\bm{\mu}^{(m)}$, the sub-problem to solve for $\mathbf{b}$ can be formulated as 
\begin{subequations}\label{eq:subproblem b}\begin{align}
&\underset{\mathbf{b}}\min~~\|\mathbf{a}+\mathbf{b}-\mathbf{1}+\bm{\mu}/\rho_2 \|^2\\
&~~\text{s.t.}~~\frac{2|\mathbf{b}^T\mathbf{h}|^2-|\mathbf{b}^T\mathbf{1}|^2}{\mathbf{b}^T(\mathbf{Q}_2-\mathbf{Q}_1)\mathbf{b}} \leq t_1,\\
&\qquad~~1\leq\mathbf{1}^T\mathbf{b}\leq N-K,\\
&\qquad~~0\leq b_n \leq 1,~\forall n.
\end{align}\end{subequations}
We see that the non-convex constraint (\ref{eq:subproblem b}b) is the major difficulty in solving this problem. 
According to basic linear algebra laws, the fractional form can be converted into a polynomial expression by multiplying  $\mathbf{b}^T(\mathbf{Q}_2-\mathbf{Q}_1)\mathbf{b}$ on both sides of the inequality.
In particular, constraint (\ref{eq:subproblem b}b) is equivalent to the combination of the following two sets of constraints: 
\begin{subequations}\label{eq:bc1}
\begin{align}
& \mathbf{b}^T(\mathbf{Q}_2-\mathbf{Q}_1)\mathbf{b} > 0,\\
&2|\mathbf{b}^T\mathbf{h}|^2-|\mathbf{b}^T\mathbf{1}|^2-t_1\mathbf{b}^T\mathbf{Q}_2\mathbf{b}+t_1 \mathbf{b}^T\mathbf{Q}_1\mathbf{b} \leq 0,
\end{align}\end{subequations}
and 
\begin{subequations}\label{eq:bc2}
\begin{align}
& \mathbf{b}^T(\mathbf{Q}_2-\mathbf{Q}_1)\mathbf{b} < 0,\\
&2|\mathbf{b}^T\mathbf{h}|^2-|\mathbf{b}^T\mathbf{1}|^2-t_1\mathbf{b}^T\mathbf{Q}_2\mathbf{b}+t_1 \mathbf{b}^T\mathbf{Q}_1\mathbf{b} \geq 0.   
\end{align}\end{subequations}
A variable $\mathbf{b}$ that satisfies constraint (\ref{eq:subproblem b}b) should also meet one of the constraints in (\ref{eq:bc1}) or (\ref{eq:bc2}). 

Both of these sets of constraints contain non-convex terms that must be tackled. 
As before, we employ a first-order Taylor expansion to construct lower-bounds for the terms $|\mathbf{b}^T\mathbf{h}|$, $|\mathbf{b}^T\mathbf{1}|$, and $\mathbf{b}^T\mathbf{Q}_i\mathbf{b},~i = 1, 2$, as
\begin{subequations}\label{eq:surro3}
\begin{align}
\label{eq:lb for b1}
|\mathbf{b}^T\mathbf{1}|^2 &= \mathbf{b}^T\mathbf{1}_N\mathbf{b}\geq 2\mathbf{b}^T\mathbf{1}_N\mathbf{b}^{(m)} - |\mathbf{1}^T\mathbf{b}^{(m)}|^2,\\
\label{eq:lb for bh2}
|\mathbf{b}^T\mathbf{h}|^2 &\geq 2\Re\{\mathbf{b}^T\mathbf{hh}^H\mathbf{b}^{(m)}\} - |\mathbf{h}^H\mathbf{b}^{(m)}|^2,\\
\label{eq: lb for bQb}
\mathbf{b}^T\mathbf{Q}_i\mathbf{b} &\geq 2\mathbf{b}^T\mathbf{Q}_i\mathbf{b}^{(m)} - (\mathbf{b}^{(m)})^T\mathbf{Q}_i\mathbf{b}^{(m)}.
\end{align}\end{subequations}
Moreover, we note that the matrices $\mathbf{Q}_1$ and $\mathbf{Q}_2$ are positive definite, while the value of $t_1$ can be either positive or negative.

When $t_1 > 0$, using the surrogate functions in (\ref{eq:surro3}), the constraints in (\ref{eq:bc1}) can be transformed to
\begin{subequations}\label{eq:bc1 surr1}
\begin{align}
&2\mathbf{b}^T\mathbf{Q}_2\mathbf{b}^{(m)} -(\mathbf{b}^{(m)})^T\mathbf{Q}_2\mathbf{b}^{(m)}-\mathbf{b}^T\mathbf{Q}_1\mathbf{b} > 0,\\
&2|\mathbf{b}^T\mathbf{h}|^2+ t_1 \mathbf{b}^T\mathbf{Q}_1\mathbf{b} - \mathbf{b}^T\mathbf{f}_1^{(m)} + c_1^{(m)} \leq 0,
\end{align}\end{subequations}
where we define $\mathbf{f}_1^{(m)}\triangleq \mathbf{2}_N\mathbf{b}^{(m)}+2t_1\mathbf{Q}_2\mathbf{b}^{(m)}$ and $c_1^{(m)} \triangleq |\mathbf{1}^T\mathbf{b}^{(m)}|^2 + t_1(\mathbf{b}^{(m)})^T\mathbf{Q}_2\mathbf{b}^{(m)}$, and constraint (\ref{eq:bc2}) becomes
\begin{subequations}\label{eq:bc2 surr1}\begin{align}
&\mathbf{b}^T\mathbf{Q}_2\mathbf{b} -2\mathbf{b}^T\mathbf{Q}_1\mathbf{b}^{(m)} + (\mathbf{b}^{(m)})^T\mathbf{Q}_1\mathbf{b}^{(m)} < 0,\\
&\mathbf{b}^T\mathbf{f}_2^{(m)} -|\mathbf{b}^T\mathbf{1}|^2-t_1\mathbf{b}^T\mathbf{Q}_2\mathbf{b} - c_2^{(m)} \geq 0,
\end{align}\end{subequations}
where we define $\mathbf{f}_2^{(m)}\triangleq 4\Re\{\mathbf{hh}^H\mathbf{b}^{(m)}+2t_1\mathbf{Q}_1\mathbf{b}^{(m)}\}$ and $c_2^{(m)}\triangleq 2|\mathbf{h}^H\mathbf{b}^{(m)}|^2 + t_1 (\mathbf{b}^{(m)})^T\mathbf{Q}_1\mathbf{b}^{(m)}$. 

When $t_1 < 0$, similar derivations are used to convert the constraints in (\ref{eq:bc1}) to
\begin{subequations}\label{eq:bc1 surr2}\begin{align}
&2\mathbf{b}^T\mathbf{Q}_2\mathbf{b}^{(m)} -(\mathbf{b}^{(m)})^T\mathbf{Q}_2\mathbf{b}^{(m)}-\mathbf{b}^T\mathbf{Q}_1\mathbf{b} > 0,\\
&2|\mathbf{b}^T\mathbf{h}|^2- t_1 \mathbf{b}^T\mathbf{Q}_2\mathbf{b} - \mathbf{b}^T\widetilde{\mathbf{f}}_1^{(m)} + \widetilde{c}_1^{(m)} \leq 0,
\end{align}\end{subequations}
where $\widetilde{\mathbf{f}}_1^{(m)}\triangleq \mathbf{2}_N\mathbf{b}^{(m)}-2t_1\mathbf{Q}_1\mathbf{b}^{(m)}$ and $\widetilde{c}_1^{(m)} \triangleq |\mathbf{1}^T\mathbf{b}^{(m)}|^2 - t_1(\mathbf{b}^{(m)})^T\mathbf{Q}_1\mathbf{b}^{(m)}$, and to convert the constraints in (\ref{eq:bc2}) to
\begin{subequations}\label{eq:bc2 surr2}\begin{align}
&\mathbf{b}^T\mathbf{Q}_2\mathbf{b} -2\mathbf{b}^T\mathbf{Q}_1\mathbf{b}^{(m)} + (\mathbf{b}^{(m)})^T\mathbf{Q}_1\mathbf{b}^{(m)} < 0,\\
&\mathbf{b}^T\widetilde{\mathbf{f}}_2^{(m)} -|\mathbf{b}^T\mathbf{1}|^2+t_1\mathbf{b}^T\mathbf{Q}_1\mathbf{b} - \widetilde{c}_2^{(m)} \geq 0,
\end{align}\end{subequations}
where we define $\widetilde{\mathbf{f}}_2^{(m)}\triangleq 4\Re\{\mathbf{hh}^H\mathbf{b}^{(m)}-2t_1\mathbf{Q}_2\mathbf{b}^{(m)}\}$ and $\widetilde{c}_2^{(m)}\triangleq 2|\mathbf{h}^H\mathbf{b}^{(m)}|^2 - t_1 (\mathbf{b}^{(m)})^T\mathbf{Q}_2\mathbf{b}^{(m)}$.

Finally, the problem for updating $\mathbf{b}$ is given by 
\begin{subequations}\label{eq:solve for b}\begin{align}
&\underset{\mathbf{b}}\min~~\|\mathbf{a}+\mathbf{b}-\mathbf{1}+\bm{\mu}/\rho_2 \|^2\\
&~~\text{s.t.}~~(\ref{eq:bc1 surr1})~\text{or}~(\ref{eq:bc2 surr1}),~\text{if}~ t_1 >0,\\
&\qquad~~(\ref{eq:bc1 surr2})~\text{or}~(\ref{eq:bc2 surr2}),~\text{if}~ t_1 < 0,\\
&\qquad~~1\leq\mathbf{1}^T\mathbf{b}\leq N-K,\\
&\qquad~~0\leq b_n \leq 1,~\forall n,
\end{align}\end{subequations}
which is a convex QCQP problem that can be easily solved.

\subsubsection{Update $\bm{\mu}$}
After obtaining the other variables, the dual variable $\bm{\mu}$ is updated by
\begin{equation}
\bm{\mu}^{(m+1)} = \bm{\mu}^{(m)} + \rho_2(\mathbf{a}^{(m+1)} + \mathbf{b}^{(m+1)}-\mathbf{1} ).
\label{eq:update mu}
\end{equation}

\subsection{Initialization}

The proposed alternating algorithm implements a local search, and its performance relies on the initial point $\{\mathbf{W}^{(0)}, \mathbf{a}^{(0)}, \mathbf{b}^{(0)}, t_1^{(0)}, t_2^{(0)},\bm{\mu}^{(0)}\}$. 
Thus, an appropriate initialization is important for solving (\ref{eq:original problem}) using the proposed algorithm.
In this subsection, we derive a simple but effective method to sequentially initialize these variables.

In order to guarantee fairness among different antennas, we assume that the initial array partitioning coefficient is $a^{(0)}_n = b^{(0)}_n = 0.5,~\forall n$, i.e., all antennas are equally likely to be selected as transmit or receive antennas.
Then, we choose the transmit beamformer $\mathbf{W}$ to maximize the power of the received target echoes as follows: 
\begin{subequations}\label{eq:problem for W0}\begin{align}
&\underset{\mathbf{W}}\max~~\|(\mathbf{I}_N-\mathbf{A})\mathbf{h}_\text{t}\mathbf{h}_\text{t}^T\mathbf{A}\mathbf{W}\|_F^2\\
&~~\text{s.t.}~~\frac{|\mathbf{h}_{k}^T\mathbf{A}\mathbf{w}_{k}|^2}
{\sum_{j\neq k}^{K+N}|\mathbf{h}_{k}^T\mathbf{A}\mathbf{w}_{j}|^2+\sigma_k^2}\geq \Gamma_k,~\forall k,\\
&\qquad~~\|\mathbf{A}\mathbf{W}\|_F^2 \leq P.
\end{align}\end{subequations}
Based on the derivations in (\ref{eq:transf num}), the objective function (\ref{eq:problem for W0}a) can be equivalently transformed to 
\be
\underset{\mathbf{W}}\max~~\mathbf{a}^T\text{diag}\{\mathbf{h}_\text{t}\}\mathbf{WW}^H\text{diag}\{\mathbf{h}_\text{t}^*\}\mathbf{a}.
\ee

It is obvious that the non-concave objective function and the constraints with quadratic fractional terms hinder a direct solution. 
In order to tackle this difficulty, we define $\mathbf{R} \triangleq \mathbf{WW}^H$ and $\mathbf{R}_k \triangleq \mathbf{w}_k\mathbf{w}_k^H,~\forall k$, to convert the quadratic terms into primary variables, and then we transform problem (\ref{eq:problem for W0}) into
\begin{subequations}\label{eq:problem for W0 with rank-one}\begin{align}
&\underset{\mathbf{R},\mathbf{R}_k,\forall k}\max
~\text{Tr}\big\{\text{diag}\{\mathbf{h}_\text{t}^*\}\mathbf{a}\mathbf{a}^T\text{diag}\{\mathbf{h}_\text{t}\}\mathbf{R}\big\}\\
&\quad\text{s.t.}\quad(1+\Gamma_k^{-1})\text{Tr}\{\mathbf{A}\mathbf{h}_k^*\mathbf{h}_k^T\mathbf{A}\mathbf{R}_k\}\non\\
&\qquad\qquad\qquad -\text{Tr}
\{\mathbf{A}\mathbf{h}_k^*\mathbf{h}_k^T\mathbf{A}\mathbf{R}\}\geq \sigma_k^2,~\forall k, \\
&\qquad\quad\text{Tr}\{\mathbf{A}\mathbf{R}\mathbf{A}\} \leq P,\\
&\qquad\quad\mathbf{R}\in\mathcal{S}_N^+,~\mathbf{R}_k\in\mathcal{S}_N^+,~\forall k,~\mathbf{R}-\sum_{k=1}^K\mathbf{R}_k\in\mathcal{S}_N^+,\\
&\qquad\quad\text{Rank}\{\mathbf{R}_k\}=1,~\forall k. \label{eq:rank-one constraint}
\end{align}\end{subequations}
This is a typical semi-definite relaxation (SDR) approach, and the problem is solved by neglecting the rank-one constraint (\ref{eq:rank-one constraint}), in which case (\ref{eq:problem for W0 with rank-one}) becomes a semi-definite programming (SDP) problem:
\begin{equation}\label{eq:SDP problem for W0}\begin{aligned}
&\underset{\mathbf{R},\mathbf{R}_k,\forall k}\max
~\text{Tr}\big\{\text{diag}\{\mathbf{h}_\text{t}^*\}\mathbf{a}\mathbf{a}^T\text{diag}\{\mathbf{h}_\text{t}\}\mathbf{R}\big\}\\
&\quad~\text{s.t.}~\quad(\ref{eq:problem for W0 with rank-one}\text{b})-(\ref{eq:problem for W0 with rank-one}\text{d}),
\end{aligned}\end{equation}
whose objective function and constraints are linear or semi-definite, and a globally optimal solution can be readily found in polynomial time using standard convex optimization tools.

We denote the solutions to problem (\ref{eq:SDP problem for W0}) as $\widetilde{\mathbf{R}}$ and $\widetilde{\mathbf{R}}_k,~\forall k$.
This solution may not be unique or satisfy the rank-one constraint.
However, as shown in \cite{Liu-TSP-2020}, there exists a globally optimal solution satisfying the rank-one constraint (\ref{eq:rank-one constraint}), and this solution can be computed using the obtained $\widetilde{\mathbf{R}}$ and $\widetilde{\mathbf{R}}_k,~\forall k$.
In particular, once $\widetilde{\mathbf{R}}$ and $\widetilde{\mathbf{R}}_k$ have been obtained from (\ref{eq:SDP problem for W0}), the optimal solutions $\mathbf{R}^\star$, $\mathbf{R}_k^\star$, and $\mathbf{w}_k^\star$ to problem (\ref{eq:problem for W0 with rank-one}) can be found as:
\begin{subequations}\label{eq:Rstar}\begin{align}
\mathbf{R}^\star &= \widetilde{\mathbf{R}}, \\
\mathbf{w}_k^\star &= (\mathbf{h}_k^T\mathbf{A}\widetilde{\mathbf{R}}_k\mathbf{A}\mathbf{h}_k^*)^{-1/2}\widetilde{\mathbf{R}}_k\mathbf{A}\mathbf{h}_k^*,~\forall k,\label{eq:initial wk}\\
\mathbf{R}_k^\star &=  \mathbf{w}_k^\star(\mathbf{w}_k^\star)^H.
\end{align}\end{subequations}
A proof for the results in (\ref{eq:Rstar}) is straightforward and follows the same procedure as in \cite{Liu-TSP-2020}, and thus we omit it here for brevity.
Recalling that $\mathbf{R} = \mathbf{WW}^H=\mathbf{W}_\text{c}\mathbf{W}_\text{c}^H+\mathbf{W}_\text{r}\mathbf{W}_\text{r}^H$, the radar beamformer $\mathbf{W}_\text{r}$ satisfies
\be\label{eq:initial Wr}
\mathbf{W}_\text{r}\mathbf{W}_\text{r}^H = \mathbf{R}^\star-\sum_{k=1}^K\mathbf{R}_k^\star,
\ee
from which we can calculate the optimal radar beamformer $\mathbf{W}_\text{r}^\star$ using a Cholesky or eigenvalue decomposition.
\begin{algorithm}[!t]
\begin{small}
\caption{Joint Array Partitioning and Transmit Beamforming Design}
\label{alg1}
    \begin{algorithmic}[1]
    \REQUIRE $\mathbf{h}_\text{t}$, $\sigma_\text{t}^2$, $\sigma_\text{r}^2$, $\mathbf{h}_k$, $\sigma_k^2$, $\Gamma_k$, $\forall k$, $\mathbf{H}_\text{SI}$, $P$, $\rho_1$, $\rho_2$, $N_\text{max}$, $\delta_\text{th}$.
    \ENSURE $\mathbf{a}^\star$, $\mathbf{W}_\text{c}^\star$, and $\mathbf{W}_\text{r}^\star$.        
        \STATE {Initialize $a^{(0)}_n=b^{(0)}_n = 0.5,~\forall n$.}
        \STATE {Obtain $\widetilde{\mathbf{R}}$ and $\widetilde{\mathbf{R}}_k,~\forall k$, by solving problem (\ref{eq:SDP problem for W0}).}
        \STATE {Calculate $\mathbf{w}_k^\star,~\forall k$, by (\ref{eq:initial wk}).}
        \STATE {Calculate $\mathbf{W}_\text{r}^\star$ using Cholesky decomposition based on (\ref{eq:initial Wr}).}
        \STATE {Construct $\mathbf{W} = [\mathbf{w}_1^\star,\mathbf{w}_2^\star,\ldots,\mathbf{w}_K^\star,\mathbf{W}_\text{r}^\star]$.}
        \STATE {Initialize $\mathbf{W}^{(0)} = \mathbf{W}$ and $\bm{\mu}^{(0)} = \mathbf{0}$.}
        \STATE {Calculate $t_1^{(0)}$ and $t_2^{(0)}$ by (\ref{eq:update t12}).}
        \STATE {Set $\rho_1 = \rho_2  = (c_1+c_2t_1^{(0)})/\sqrt{t_2^{(0)}}$.}
        \STATE {Set $m = 0$ and $\delta = \infty$.}
        \STATE{Calculate the objective value of (\ref{eq:problem after alpha}a) as $f^{(m)}$.}
        \WHILE {$m < N_\text{max}$ and $\delta > \delta_\text{th}$ }        	
            \STATE{Update $\mathbf{W}^{(m+1)}$ by solving (\ref{eq:solve for W}).}
            \STATE{Update $\mathbf{a}^{(m+1)}$ by solving (\ref{eq:solve for a}).}
            \STATE{Update $\mathbf{b}^{(m+1)}$ by solving (\ref{eq:solve for b}).}
            \STATE{Update $\bm{\mu}^{(m+1)}$ by (\ref{eq:update mu}).}            
            \STATE{Update $t_1^{(m+1)}$ and $t_2^{(m+1)}$ by (\ref{eq:update t12}).}
            \STATE{Calculate the objective value of (\ref{eq:problem after alpha}a) as $f^{(m+1)}$.}
            \STATE{Calculate $\delta = |1-f^{(m+1)}/f^{(m)}|$.}
            \STATE{$m := m+1$.}
        \ENDWHILE
        \STATE{Refine $\mathbf{a}$ using (\ref{eq:refine a}) and re-design $\mathbf{W}$.}
        \STATE{Return $\mathbf{a}^\star=\mathbf{a}$, $\mathbf{W}_\text{c}^\star = \mathbf{W}(:,1:K)$, $\mathbf{W}_\text{r}^\star = \mathbf{W}(:,K+1:\text{end})$.}
    \end{algorithmic}
    \end{small}
\end{algorithm}

\subsection{Summary and Complexity Analysis}

The proposed joint array partitioning and transmit beamforming design algorithm is summarized in Algorithm 1, where $N_\text{max}$ denotes the maximum allowable number of iterations and $\delta_\text{th}$ is the threshold to determine convergence. 
The parameters $\mathbf{a}^{(0)}$, $\mathbf{b}^{(0)}$, $\mathbf{W}^{(0)}$, $t_1^{(0)}$, $t_2^{(0)}$, and $\bm{\mu}^{(0)}$ are first sequentially initialized.
The penalty parameters $\rho_1$ and $\rho_2$ are set equal to the initial objective value, i.e., $\rho_1=\rho_2=(c_1+c_2t_1^{(0)})/\sqrt{t_2^{(0)}}$, to ensure they are appropriately scaled relative to the objective, balancing convergence speed and achievable performance. Then, the transmit beamformer $\mathbf{W}$, the array partitioning vector $\mathbf{a}$, the auxiliary variable $\mathbf{b}$, the dual variable $\bm{\mu}$, and the auxiliary variables $t_1$ and $t_2$ are iteratively updated using (\ref{eq:solve for W}), (\ref{eq:solve for a}), (\ref{eq:solve for b}), (\ref{eq:update mu}), and (\ref{eq:update t12}) respectively, until the relative difference of the objective value is less than the pre-set threshold. Since the penalty parameter $\rho_1$ may not be large enough to yield a $\mathbf{a}$ that exactly satisfies the binary integer constraints, we round the elements of the obtained solution for $\mathbf{a}$ after convergence to ensure they are either $0$ or $1$:
\be\label{eq:refine a}
\mathbf{a} = \mathrm{round}\left[\mathbf{a}\right].
\ee
Given this refined array partitioning vector, the transmit beamformer $\mathbf{W}$ is re-designed using the algorithm presented in the next section.

The primary computational cost of Algorithm 1 is due to the QCQP problems (\ref{eq:solve for W}), (\ref{eq:solve for a}) and (\ref{eq:solve for b}) for updating $\mathbf{W}$, $\mathbf{a}$, and $\mathbf{b}$, respectively.
Assuming that the general interior-point method is used to solve these problems, 
the computational complexity of solving problem (\ref{eq:solve for W}) with an $N(N+K)$-dimensional variable, problem (\ref{eq:solve for a}) with an $N$-dimensional variable, and problem (\ref{eq:solve for b}) with an $N$-dimensional variable is of order $\mathcal{O}\big\{N^{3.5}(N+K)^{3.5}\big\}$, $\mathcal{O}\big\{N^{3.5}\big\}$, and $\mathcal{O}\big\{N^{3.5}\big\}$, respectively. 
Thus, the computational complexity for one iteration of Algorithm 1 is of order $\mathcal{O}\big\{N^{3.5}(N+K)^{3.5}\}$. 
Since the proposed algorithm requires an alternating optimization, and each iteration requires the update of six variables, the overall complexity of Algorithm~1 could be large, depending on the required number of iterations. To provide a balance between performance and computational complexity, in the next section we will develop a more efficient heuristic approach to solve the considered problem.

\section{Heuristic Joint Array Partitioning and Transmit Beamforming Design}\label{sec:Algorithm 2}

In this section, we present a heuristic algorithm to significantly decrease the computational complexity of the solution to the joint array partitioning and transmit beamforming design problem. The discrete nature of the array partitioning vector $\mathbf{a}$ is the primary cause of the complexity in the original algorithm due to three transformations required to handle it. The first transforms the binary integer constraint for each element of $\mathbf{a}$ into a continuous constraint with a quadratic penalty term; the second converts two complicated constraints with respect to $\mathbf{a}$ into separate constraints related to an auxiliary variable $\mathbf{b}$ that is associated with the dual variable $\bm{\mu}$; the third transforms the non-concave terms associated with $\mathbf{a}$ and $\mathbf{b}$ into linear terms. These transformations require additional iterations to approach the local optimal solution. Thus, to reduce the complexity of the joint design of $\mathbf{a}$ and $\mathbf{W}$, we propose to split the original problem into separate designs for array partitioning and transmit beamforming.

Specifically, we first develop a straightforward yet effective array partitioning strategy and then optimize the transmit beamforming for the resulting DOA estimation-oriented design. Since the DOA estimation performance is related to both the number of receive antennas and the resulting array aperture, we further convert the design of $\mathbf{a}$ into finding the optimal value of $N_\text{r}/N_\text{t}$ and then separately determine the index sets for the transmit/receive antennas. Thus, a three-step strategy is proposed to design the antenna array partitioning and transmit beamforming, as described below.

\textbf{Step 1: Calculate the number of transmit/receive antennas.} To achieve better DOA estimation performance and simplify the optimization, we calculate the number of receive antennas based on an ideal radar sensing scenario where the transmit beamformer is aligned with the channel of the target to guarantee satisfactory sensing performance. Under this assumption, the covariance of the transmit beamformer will be 
\begin{equation}\label{eq:radar R}
	\mathbf{WW}^H = \frac{P}{\beta_\text{t}^2N}\mathbf{h}_\text{t}^*\mathbf{h}_\text{t}^T.
\end{equation}
Based on this, we formulate the DOA estimation performance with respect to the number of receive antennas $N_\text{r}$. 

As shown in (\ref{eq:sigma theta}), the DOA RMSE is determined by the 3dB beamwidth and the receive SINR. For the considered array partitioning scenario, the 3dB beamwidth is upper-bounded by $\theta_\text{3dB} \leq 1.772/N_\text{r}$. 
Thus, we have
\be\label{eq: ub RMSE}
\sigma_\theta \leq \frac{1.772}{1.6N_\text{r}\sqrt{2\text{SINR}_\text{r}}} = \sqrt{\frac{0.6133}{N_\text{r}^2\text{SINR}_\text{r}}}.
\ee
Using the derivations in Sec. III-A, a lower bound for $\text{SINR}_\text{r}$ is obtained as in (\ref{eq:SINR lb}). 
Then, based on (\ref{eq:radar R}), we have the following equivalent transformations
\begin{subequations}\begin{align}
&\mathbf{a}^T
	\text{diag}\{\mathbf{h}_\text{t}\}\mathbf{WW}^H\text{diag}\{\mathbf{h}_\text{t}^H\}\mathbf{a}\nonumber\\
	& = \frac{P}{\beta_\text{t}^2N}\|\mathbf{a}^T\text{diag}\{\mathbf{h}_\text{t}\}\mathbf{h}_\text{t}^*\|_2^2\\
 &= \frac{P}{\beta_\text{t}^2N}\|\mathbf{a}^T\mathbf{1}\|_2^2=\frac{P(N-N_\text{r})^2}{\beta_\text{t}^2N}.\label{eq:trans1}
\end{align}\end{subequations}
In addition, an upper bound for the term $\|\mathbf{H}_\text{SI}\mathbf{A}\mathbf{W}\|_F^2$ is constructed as 
\begin{subequations}\begin{align}
\|\mathbf{H}_\text{SI}\mathbf{A}\mathbf{W}\|_F^2  &= \frac{P}{\beta_\text{t}^2N}
\text{Tr}\big\{\mathbf{H}_\text{SI}\mathbf{A}\mathbf{h}_\text{t}^*\mathbf{h}_\text{t}^T\mathbf{A}\mathbf{H}_\text{SI}^H\big\}\\
& = \frac{P}{\beta_\text{t}^2N}\mathbf{a}^T\text{diag}\{\mathbf{h}_\text{t}\}\mathbf{H}_\text{SI}^H\mathbf{H}_\text{SI}\text{diag}\{\mathbf{h}_\text{t}^*\}\mathbf{a}\\
&\leq \frac{P}{\beta_\text{t}^2N}\lambda_\text{m}\mathbf{a}^T\mathbf{a}= \frac{P}{\beta_\text{t}^2N}\lambda_\text{m}(N-N_\text{r}),\label{eq:trans2}
\end{align}\end{subequations}
where $\lambda_\text{m} > 0$ is the largest eigenvalue of the matrix $\text{diag}\{\mathbf{h}_\text{t}\}\mathbf{H}_\text{SI}^H\mathbf{H}_\text{SI}\text{diag}\{\mathbf{h}_\text{t}^*\}$. 
Substituting the results of (\ref{eq:trans1}), (\ref{eq:trans2}) and (\ref{eq:SINR lb}) into the objective function (\ref{eq: ub RMSE}), an upper bound for $\sigma_\theta^2$ is obtained as 
\begin{equation}
        \sigma_\theta^2 \leq
 \frac{0.6133(\beta_\text{t}^2N\sigma_\text{r}^2 + P\lambda_\text{m}(N-N_\text{r}))}
 {\beta_\text{t}^2\sigma_\text{t}^2PN_\text{r}^2(N-N_\text{r})^2}.
\end{equation}
Thus, the optimization for $N_\text{r}$ can be formulated as 
\begin{subequations}\label{eq:solve for Nt}\begin{align}
&\underset{N_\text{r}}\min
~~\frac{\beta_\text{t}^2N\sigma_\text{r}^2 + P\lambda_\text{m}(N-N_\text{r})}{\beta_\text{t}^2\sigma_\text{t}^2PN_\text{r}^2(N-N_\text{r})^2}\\
&\quad\text{s.t.}~~1\leq N_\text{r} \leq N-K,
\end{align}\end{subequations}
which is a relatively simple problem with only one unknown variable. 

A one-dimensional line search can be used to find the optimal solution $N_\text{r}^\star$. 
Since this solution is derived for the radar-only scenario with a worst-case antenna distribution, the obtained value of $N_\text{r}^\star$ will provide good sensing results, but it does not consider the multiuser communication performance.
To remedy this issue, we allocate $K$ elements from the receive antenna set to the transmit antenna set so that the number of receive and transmit antennas are respectively $N_\text{r} = N_\text{r}^\star-K$ and $N_\text{t} = N - N_\text{r}$.

\textbf{Step 2: Determine the indices of the transmit/receive antennas.} After calculating the number of transmit/receive antennas based on the ideal radar sensing scenario, we propose to determine their indices (locations) from the multiuser communications perspective. Since the communications performance relies only on the transmit antennas, we select $N_\text{t}$ antennas from the array by evaluating each antenna's impact in terms of the transmit power allocated to it. In particular, we first design the transmit beamformer for the MU-MISO communication system by solving the following optimization problem
\begin{subequations}\label{eq:power minimization}
\begin{align}
&\underset{\mathbf{W}}{\min}~~\|\mathbf{W}\|_F^2 \\
&\text{s.t.}~~\frac{|\mathbf{h}_{k}^T\mathbf{w}_{k}|^2}
{\sum_{j\neq k}^{K+N}|\mathbf{h}_{k}^T\mathbf{w}_{j}|^2+\sigma_k^2} \geq \Gamma_k,~~\forall k,
\end{align}
\end{subequations}
which can be converted into an equivalent second-order cone programming (SOCP) problem whose solution can be found by existing algorithms. The transmit power allocated to the $n$-th antenna by the optimal solution $\widetilde{\mathbf{W}}$ to (\ref{eq:power minimization}) is given by 
\begin{equation}\label{eq:pn}
    p_n = \|\widetilde{\mathbf{W}}(n,:)\|_2^2,~\forall n.
\end{equation}
We then select the $N_\text{t}$ antennas with the highest powers as the transmit antennas, and the remaining antennas will make up the set of receive antennas. 
 
\textbf{Step 3: Optimize the transmit beamforming $\mathbf{W}$.} After determining the array partitioning, we turn to optimizing the transmit beamforming matrix $\mathbf{W}$. According to the transformations presented in Sec. III-A, the optimization problem for finding $\mathbf{W}$ can be equivalently formulated as 
\begin{subequations}\begin{align}
&\underset{\mathbf{W}}\max~~\frac{\|(\mathbf{I}-\mathbf{A})\mathbf{h}_\text{t}\mathbf{h}_\text{t}^T\mathbf{AW}\|_F^2}{\sigma_\text{r}^2N_\text{r}+\|(\mathbf{I}-\mathbf{A})\mathbf{H}_\text{SI}\mathbf{AW}\|_F^2}\\
&\quad\text{s.t.}~~\text{SINR}_k \geq \Gamma_k,~\forall k,\\
&\qquad\quad \|\mathbf{AW}\|_F^2 \leq P.
\end{align}\end{subequations}
Following a similar procedure, we introduce an auxiliary variable $\gamma$ to transform the fractional objective function into a polynomial one as 
\begin{equation}\label{eq:poly obj}
\underset{\mathbf{W}}\max~~\|(\mathbf{I}-\mathbf{A})\mathbf{h}_\text{t}\mathbf{h}_\text{t}^T\mathbf{AW}\|_F^2-\gamma\|(\mathbf{I}-\mathbf{A})\mathbf{H}_\text{SI}\mathbf{AW}\|_F^2,
\end{equation}
with the optimal solution of $\gamma$ given by
\begin{equation}\label{eq:gamma tilde}
    \gamma^\star = \frac{\|(\mathbf{I}-\mathbf{A})\mathbf{h}_\text{t}\mathbf{h}_\text{t}^T\mathbf{AW}\|_F^2}{\sigma_\text{r}^2N_\text{r}+\|(\mathbf{I}-\mathbf{A})\mathbf{H}_\text{SI}\mathbf{AW}\|_F^2}.
\end{equation}
Then, to tackle the non-concave objective (\ref{eq:poly obj}), we employ the MM method to construct a linear surrogate function 
\begin{equation}
    \|\mathbf{TW}\|_F^2 \geq\! 2\Re\{\text{Tr}\{(\mathbf{W}^{(m)})^H\mathbf{T}^H\mathbf{T}\mathbf{W}\}\}\! -\! \|\mathbf{T}\mathbf{W}^{(m)}\|_F^2,
\end{equation}
where we define $\mathbf{T} \triangleq (\mathbf{I}-\mathbf{A})\mathbf{h}_\text{t}\mathbf{h}_\text{t}^T\mathbf{A}$. 
Meanwhile, the result in (\ref{eq:hwk surr}) is utilized to convexify the communication SINR constraint. 
Therefore, the optimization problem for updating $\mathbf{W}$ in the $m$-th iteration is given by
\begin{subequations}\label{eq: update W heu}\begin{align}
&\underset{\mathbf{W}}\max~~2\Re\{\text{Tr}\{\mathbf{T}^{(m)}\mathbf{W}\}\}-\gamma\|(\mathbf{I}-\mathbf{A})\mathbf{H}_\text{SI}\mathbf{AW}\|_F^2  \\
&~\text{s.t.}~~2\Re\{\mathbf{h}_k^T\mathbf{Aw}_k\mathbf{e}_k^{(m)}\} - \|\mathbf{e}_k^{(m)}\|^2 \nonumber \\
&\qquad\quad-\Gamma_k\sum_{j\neq k}^{K+N}|\mathbf{h}_{k}^T\mathbf{A}\mathbf{w}_{j}|^2-\Gamma_k\sigma_k^2 \geq 0,~\forall k,\\
&\qquad~\|\mathbf{AW}\|_F^2 \leq P,
\end{align}\end{subequations}
where $\mathbf{T}^{(m)}\triangleq(\mathbf{W}^{(m)})^H\mathbf{T}^H\mathbf{T}$.
This is a convex QCQP problem that can be easily solved.

\begin{algorithm}[!t]
\begin{small}
\caption{Heuristic Array Partitioning and Transmit Beamforming Algorithm}
\label{alg2}
   \begin{algorithmic}[1]
   \REQUIRE $\mathbf{h}_\text{t}$, $\sigma_\text{t}^2$, $\sigma_\text{r}^2$, $\mathbf{h}_k$, $\sigma_k^2$, $\Gamma_k$, $\forall k$, $\mathbf{H}_\text{SI}$, $P$, $N_\text{max}$, $\delta_\text{th}$.
   \ENSURE $\mathbf{a}^\star$, $\mathbf{W}_\text{c}^\star$, and $\mathbf{W}_\text{r}^\star$.
       \STATE {Obtain $N_\text{r}^\star$ by solving (\ref{eq:solve for Nt}).}
       \STATE {Calculate $N_\text{r} = N_\text{r}^\star-K$ and $N_\text{t} = N - N_\text{r}$.}
       \STATE{Solve problem (\ref{eq:power minimization}) to obtain $\widetilde{\mathbf{W}}$. }
       \STATE{Calculate $p_n,~\forall n$ by (\ref{eq:pn}). }
       \STATE{Select the $N_\text{t}$ antennas with the highest powers as the transmit antennas and set the array partition.}
       \STATE {Set $m=0$ and $\delta = \infty$.}
       \STATE {Calculate $f^{(0)} = \|\mathbf{TW}\|_F^2-\gamma\|(\mathbf{I}-\mathbf{A})\mathbf{H}_\text{SI}\mathbf{AW}\|_F^2$.}
       \WHILE {$m < N_\text{max}$ and $\delta > \delta_\text{th}$}
        \STATE {Update $\gamma$ by (\ref{eq:gamma tilde}).}
        \STATE {Update $\mathbf{W}$ by solving problem (\ref{eq: update W heu}).}
        \STATE {Calculate $f^{(m+1)} = \|\mathbf{TW}\|_F^2-\gamma\|(\mathbf{I}-\mathbf{A})\mathbf{H}_\text{SI}\mathbf{AW}\|_F^2$.}
        \STATE {$\delta := |1-f^{(m+1)}/f^{(m)}|$.}
        \STATE{$m := m+1$.}
        \ENDWHILE
       \STATE{Return $\mathbf{a}^\star=\mathbf{a}$, $\mathbf{W}_\text{c}^\star = \mathbf{W}(:,1:K)$, $\mathbf{W}_\text{r}^\star = \mathbf{W}(:,K+1:\text{end})$.}
   \end{algorithmic}
   \end{small}
\end{algorithm}

The resulting heuristic antenna array partitioning and beamforming design algorithm is straightforward and summarized in Algorithm 2 based on the above derivations. We first calculate the number of transmit/receive antennas using the ideal transmit beamformer for radar sensing, then determine the index sets of the transmit/receive antennas according to the power allocation for multiuser communications, and finally we optimize the integrated beamforming given the array partitioning vector. 
The primary factor that determines the computational complexity of the heuristic approach is the update for $\mathbf{W}$, which has a complexity of order $\mathcal{O}\{N^{3.5}(N+K)^{3.5}\}$. More importantly, as will be shown in simulations, Algorithm 2 converges in considerably fewer iterations than Algorithm 1. The faster convergence can be attributed to the fact that only two of the $(N(N+K)+1)$ parameters require an alternating update, while six variables out of $(N(N+K)+3N+2)$ parameters are iteratively computed in Algorithm 1. 
Thus, the computational complexity of the heuristic Algorithm 2 is significantly lower than that of Algorithm 1.
In addition, we emphasize that the theoretical complexity analysis is a worst-case upper bound for solving QCQP problems. In practice, the computational burden can be significantly reduced by leveraging problem structure, efficient initializations, approximation/decomposition methods, and hardware acceleration.

\section{Simulation Results}\label{sec:simulations}

In this section, we provide extensive simulation results to verify the advantages of joint antenna array partition and beamforming design for ISAC systems, and the effectiveness of the proposed algorithms.
The following settings are used throughout the simulations unless otherwise specified: $N=30$, $K=6$, $P=6$W, $\sigma_\text{t}^2 = 1$, $\theta = \pi/6$, $\sigma_k^2 =\sigma_\text{r}^2 = -80\text{dBm},~\forall k$, and $\Gamma=\Gamma_k = 10$dB, $\forall k$. 
We adopt a Rician fading channel model for the communication users, i.e., 
\be
\mathbf{h}_k = \sqrt{\frac{\kappa}{\kappa+1}}\mathbf{h}_k^{\text{LoS}} + \sqrt{\frac{1}{\kappa+1}}\mathbf{h}_k^{\text{NLoS}},
\ee
where $\kappa=3$dB is the Rician factor, $\mathbf{h}_k^{\text{LoS}}\triangleq [1~e^{-\jmath\pi\sin\theta_k}~\ldots~e^{-\jmath(N-1)\pi\sin\theta_k}]^T$ denotes the LoS component of the channel, $\theta_k$ is the azimuth angle of the $k$-th user with respect to the BS, and $\mathbf{h}_k^{\text{NLoS}}$ denotes the Rayleigh fading components with zero mean and unit variance. 
We assume that the azimuth angles of the users are randomly generated in $\theta_k\sim[-\pi/2,\pi/2)$.
In addition, we adopt the following typical distance-dependent path-loss model: $\text{PL}(d) = C_0(d/d_0)^{-\alpha}$, where $C_0 = -30$dB, $d_0=1$m, $d$ represents the link distance, and $\alpha$ denotes the path-loss exponent. 
We set the distances of the BS-target link and the BS-user link as $d_\text{t} = 30$m and $d_k = 50$m, $\forall k$, and the path-loss exponents for these channels are $2.8$ and $3.5$, respectively. 
Thanks to various advanced analog- and digital-domain cancellation techniques developed for full-duplex systems \cite{SI cancel 1}-\cite{Erdem 2021 WC}, self-interference can be effectively suppressed, leaving a controllable residual interference component. The residual SI channel is modeled as \cite{He JSAC 2023}: $\mathbf{H}_\text{SI}(i,j) = {\alpha_\text{SI}}e^{-\jmath 2\pi d_{i,j}/\lambda}$, where $\alpha_\text{SI}$ denotes the amplitude of the residual SI, $d_{i,j}$ is the distance between the $i$-th transmit antenna and the $j$-th receive antenna, and $\lambda$ is the wavelength. We set the residual SI power as $-60$dBm, or equivalently $\alpha_\text{SI} = 10^{-4.5}$. We set the maximum number of iterations as $N_\text{max} = 1000$ and the convergence threshold as $\delta_\text{th} = 0.001$.

\begin{figure}[!t]
\centering
\subfigure[Proposed Algorithm 1.]{
\begin{minipage}{0.22\textwidth}
\centering\label{fig:conv1}
\includegraphics[width = \linewidth]{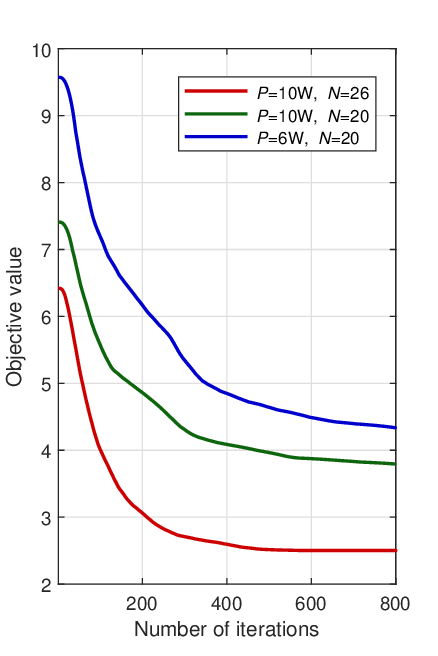}\vspace{0.1 cm}
\end{minipage}
}
\subfigure[Proposed Algorithm 2.]{
\begin{minipage}{0.22\textwidth}
\centering\label{fig:conv2}
\includegraphics[width = \linewidth]{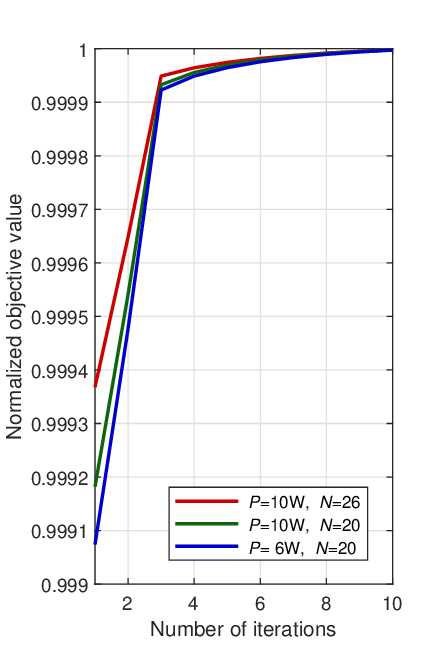}\vspace{0.1 cm}
\end{minipage}
}\vspace{-0.1 cm}
\caption{Convergence performance of the proposed algorithms.}\label{fig:convergence}\vspace{-0.4 cm}
\end{figure}

We first illustrate the convergence performance of the proposed algorithms in Fig. \ref{fig:convergence}, where the objective value of (\ref{eq:problem after alpha}a) associated with the proposed ADMM-MM-based Algorithm 1 and the normalized objective value of (\ref{eq:poly obj}) associated with the proposed heuristic Algorithm 2 are plotted versus the number of iterations. 
We observe that the proposed algorithms always converge in a limited number of iterations under different settings. While Algorithm 1 requires hundreds of iterations to reach convergence, the heuristic Algorithm 2 quickly converges within 10 iterations. 
We recall that the major computational complexity of these two algorithms arises from solving convex QCQP problems in each iteration; Algorithm 2 handles only one such problem while Algorithm 1 deals with three.
Thus, the overall computational complexity of the heuristic algorithm is dramatically lower than that of the ADMM-MM-based approach.

In order to demonstrate the advantages of the proposed ADMM-MM-based algorithm (denoted as ``\textbf{Prop., Alg. 1}'') and the proposed heuristic design algorithm (denoted as ``\textbf{Prop., Alg. 2}''), we also include the following three array partitioning schemes for comparison. The first is the ordinary array partitioning strategy that evenly divides the array into equally-sized transmit and receive subarrays, i.e., $\mathbf{a} = [\underbrace{1~1~\ldots 1}_{N/2}~\underbrace{0~0~\ldots 0}_{N/2}]^T$, denoted as ``\textbf{Even}''. To demonstrate the performance gain resulting from optimizing the specific locations of the transmit/receive antennas, we use the same number of transmit antennas obtained from Algorithm~1, $N_\text{t}$, and then partition the array into two contiguous subarrays based on this number, i.e., $\mathbf{a} = [\underbrace{1~1~\ldots 1}_{N_\text{t}}~\underbrace{0~0~\ldots 0}_{N-N_\text{t}}]^T$, referred to as ``\textbf{Cont.}'', and we also consider a random partition of the $N$ antennas into an $N_\text{t}$-element transmit array and an $N-N_\text{t}$ element receive array, referred to as ``\textbf{Rand.}''.
After applying these different strategies to determine the array partitioning vector $\mathbf{a}$, the corresponding transmit beamforming is optimized using steps $6-14$ in Algorithm 2.
The ``Even'' partition approach has the same order of complexity as Algorithm 2, while the ``Cont.'' and ``Rand.'' partitions have the highest computational complexity since they require the result from Algorithm 1 and then optimize the transmit beamforming using Algorithm 2.

\begin{figure}[t]	
	\includegraphics[width = 0.9\linewidth]{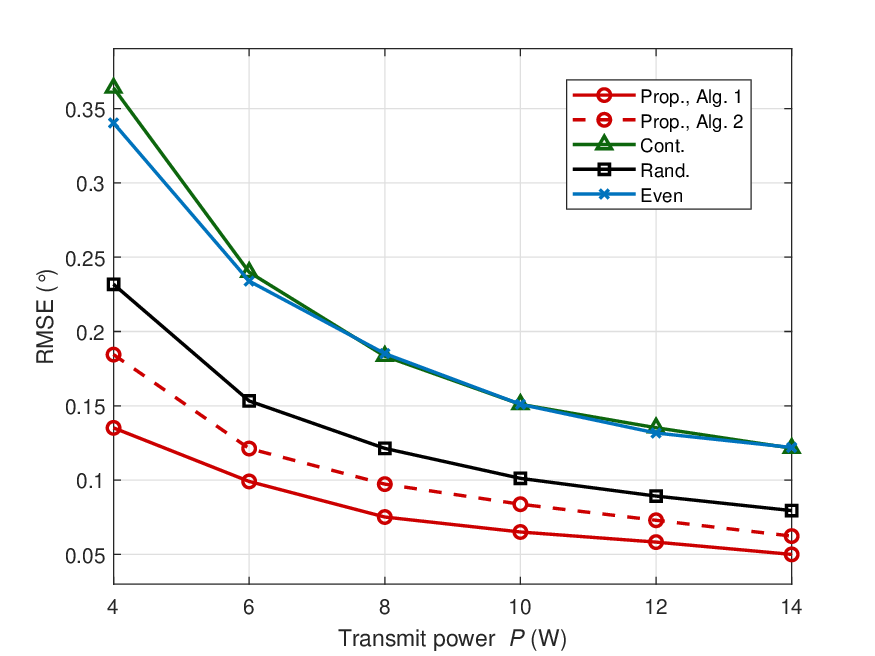}	
	\caption{DOA RMSE versus power budget $P$ for $N=30$.}\label{fig:RMSE_P}\vspace{-0.4 cm}
\end{figure}

\begin{figure}[!t]	
\centering
\subfigure[DOA RMSE]{\includegraphics[width = 0.45\linewidth]{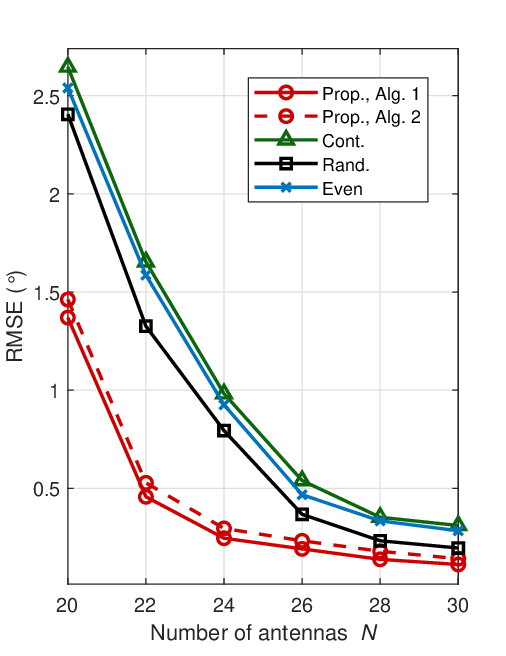}\label{fig:RMSE_N}}
\subfigure[Number of receive antennas.]{\includegraphics[width = 0.45\linewidth]{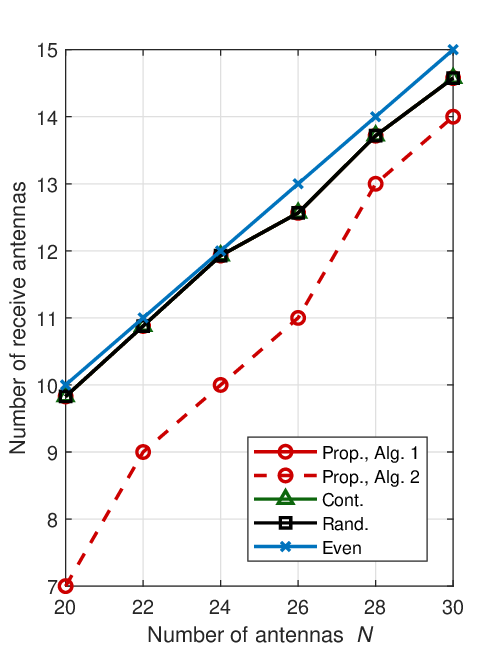}\label{fig:Nt_N}}
\caption{Performance versus number of antennas $N$ for $P=6$W.}
\vspace{-0.4 cm}
\end{figure}

Once the array partitioning and transmit beamformer are determined, we use the classical MUltiple SIgnal Classification (MUSIC) algorithm to obtain the DOA estimate $\widetilde{\theta}$, and then evaluate the DOA estimation performance by the RMSE defined as $\text{RMSE} = \sqrt{\mathbb{E}\{|\widetilde{\theta}-\theta|^2\}}$. 
The DOA RMSE is plotted versus the transmit power budget $P$ in Fig. \ref{fig:RMSE_P}. 
It is clear that a larger transmit power provides better radar sensing performance for the same communication requirements. 
We further see that the DOA estimation performance of the proposed algorithms is significantly better than the simple ``Even'' partition and ``Cont.'' partition approaches, achieving a reduction in RMSE of about 60\% and 45\% for Algorithm 1 and Algorithm 2, respectively. The ``Rand. '' partition scheme performs reasonably well in this scenario, but still with a sizable gap compared to the proposed algorithms.

\begin{figure}[t]
    \centering
    \subfigure[Proposed Algorithm 1]{
    \includegraphics[width = 0.45\linewidth]{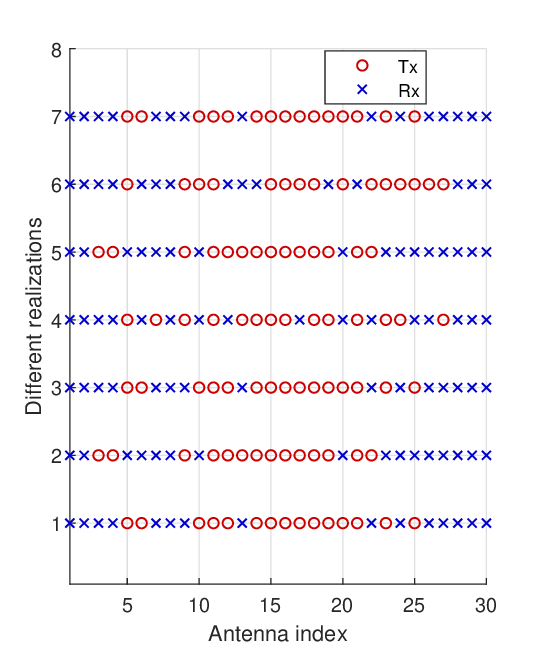}}
    \subfigure[Proposed Algorithm 2]{
    \includegraphics[width = 0.45\linewidth]{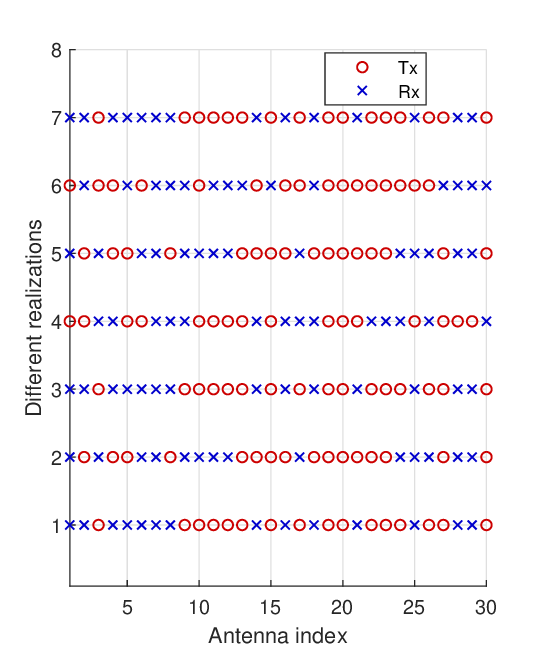} }
    \vspace{-0.1 cm}
    \caption{Examples of array partitioning for $P=6$W, $N=30$.}\vspace{-0.4 cm}
    \label{fig:array_partition}
\end{figure}

\begin{figure}[t]
    \centering
    \includegraphics[width = 0.9\linewidth]{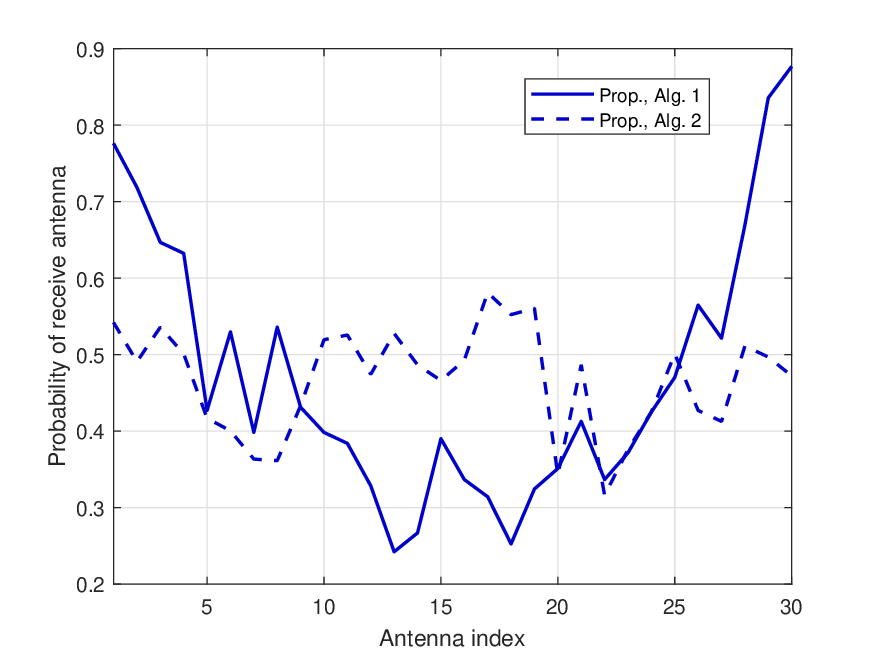}
    \caption{Probability of receive antenna locations for $P=6$W, $N=30$.}\vspace{-0.4 cm}
    \label{fig:prob}
\end{figure}

\begin{figure}[t]	
\centering
\includegraphics[width = 0.9\linewidth]{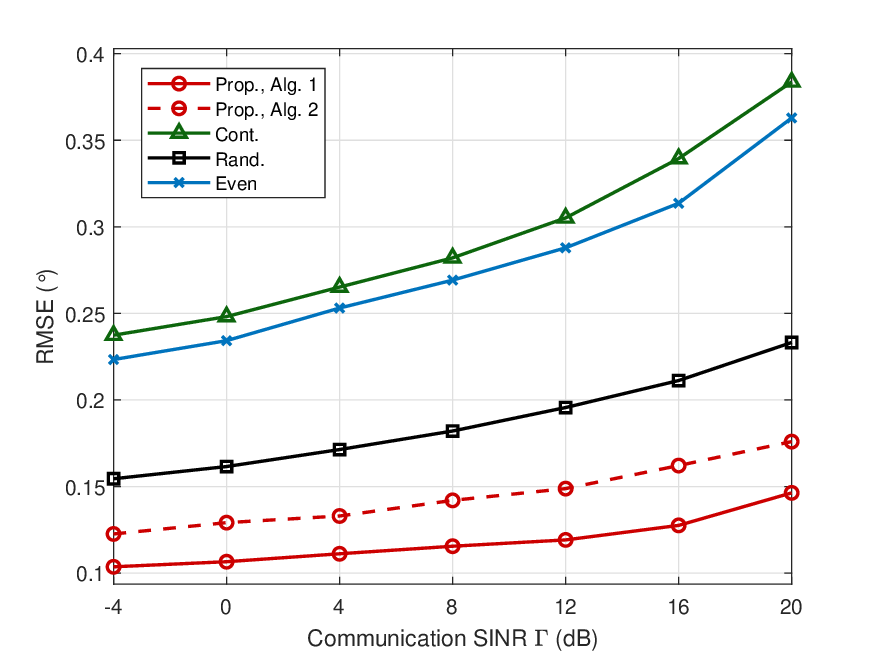}
\caption{DOA RMSE versus communication SINR $\Gamma$.}\label{fig:RMSE_SINR}\vspace{-0.4 cm}
\end{figure}

In Fig. \ref{fig:RMSE_N}, we present the DOA RMSE versus the number of antennas $N$. 
Since more antennas provide increased DoFs that can be used to boost the transmit beamforming gains and spatial signal processing capability, the RMSE decreases quickly as $N$ increases. Compared with Fig.~\ref{fig:RMSE_P}, we see that the gap in performance between the proposed algorithms and the other benchmarks increases significantly for smaller $N$, and that relatively speaking, there is little difference in the RMSE obtained by Algorithms~1 and~2. However, as $N$ grows large, the importance of the optimal partitioning diminishes.

The number of receive antennas after optimization versus the total number of antennas $N$ is illustrated in Fig. \ref{fig:Nt_N}. Note that the antenna allocation for Algorithm~1 follows the same curve as the ``Cont.'' and ``Rand.'' partitions, and thus is not explicitly visible. Interestingly, in this case the optimal number of receive antennas found by Algorithm 1 is always close to $N/2$, so its improved performance compared with the other benchmarks is due to the optimized partitioning of the array into transmit and receive antennas. We also observe that the heuristic approach in Algorithm~2 employs fewer receive antennas than the other approaches. This is because Algorithm~2 allocates an additional $K$ elements to the transmit antenna set from the ideal receive antenna set for radar sensing in order to guarantee the communication requirements.

To illustrate the spatial distribution of the transmit and receive antennas after optimization, we plot several examples of the array partitioning in Fig.~\ref{fig:array_partition}.
We see that both Algorithm 1 and 2 yield a partitioning with widely distributed receive antennas, since this provides a larger effective aperture and can exploit spatial sparsity to improve DOA estimation performance with fewer antennas, allowing more antennas to be used to enhance the transmit beamforming gains. In addition, in 
Fig.~\ref{fig:prob} we plot the probability of each possible antenna position being allocated as a receive antenna based on an average over 500 channel realizations. Compared with Algorithm 2, Algorithm 1 tends to push the receive antennas to the ends of the array to create the largest possible array aperture, which likely explains its improved DOA estimation performance.

\begin{figure}[t]
\centering
\includegraphics[width = 0.9\linewidth]{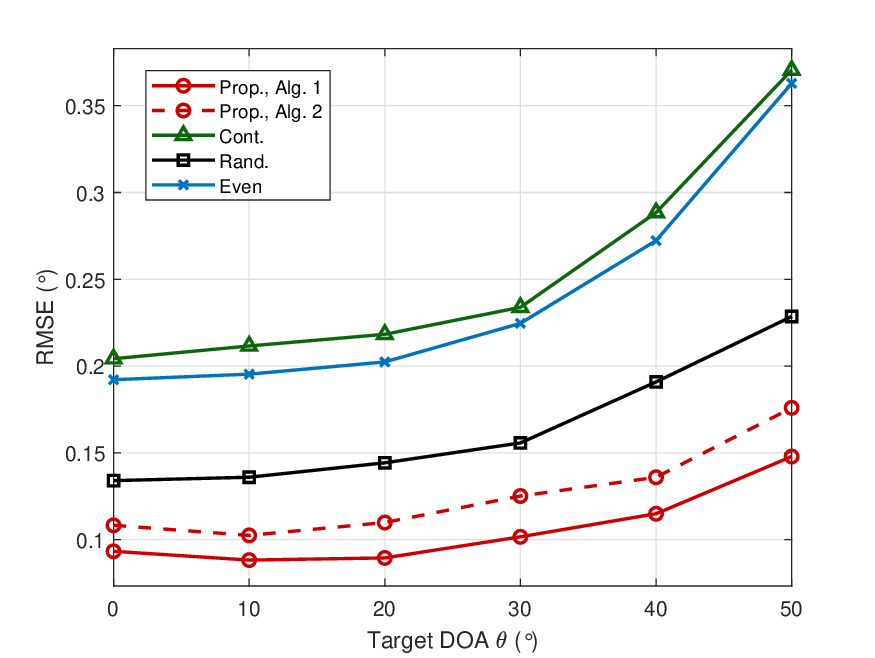}
\caption{DOA RMSE versus the target DOA.}\label{fig:DOA}\vspace{-0.3 cm}
\end{figure}

Next, the DOA RMSE versus the communication SINR requirement $\Gamma$ is illustrated in Fig. \ref{fig:RMSE_SINR}. The performance trade-off between multiuser communications and target DOA estimation is clearly observed, as the RMSE of all approaches grows as $\Gamma$ increases. 
Finally, the impact of the target DOA is demonstrated in Fig. \ref{fig:DOA}. A larger RMSE is observed as the target DOA moves away from the array broadside, which is typical for linear arrays. The key take-away from this plot is that the proposed algorithms are less sensitive to the value of the target DOA since the transmit/receive antenna locations can be optimized in favor of more oblique angles of arrival.

\begin{figure}[!t]	
\centering
\subfigure[SI/target echo channel power ratio.]{\includegraphics[width = 0.49\linewidth]{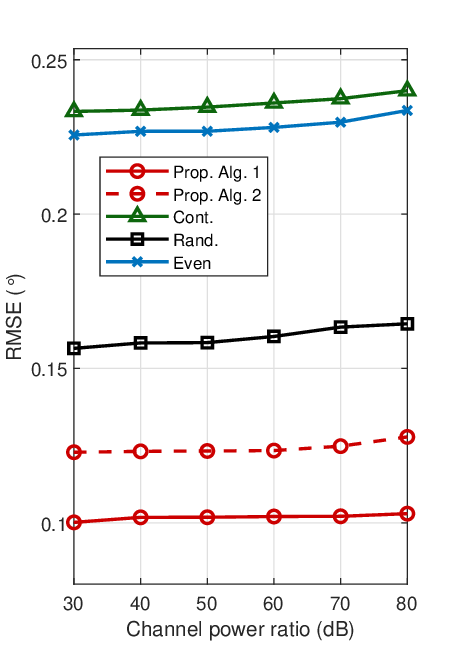}\label{fig:SI}}
\subfigure[SI channel uncertainty.]{\includegraphics[width = 0.49\linewidth]{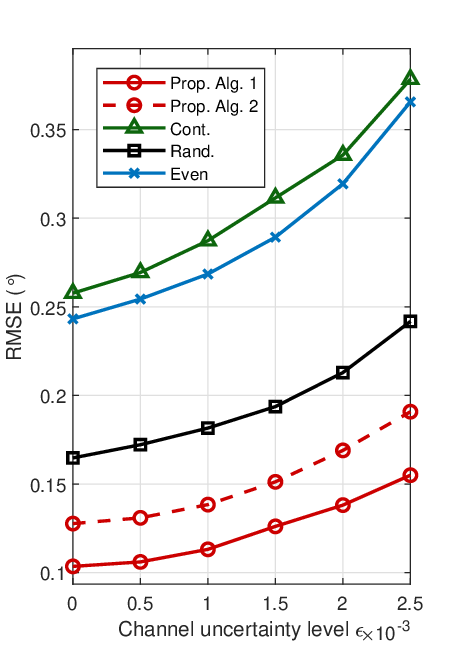}\label{fig:ISI}}
\caption{RMSE versus SI channel conditions.}
\label{fig:SI channel}\vspace{-0.3 cm}
\end{figure}

Finally, we evaluate the impact of SI channel conditions in Fig. \ref{fig:SI channel}. Fig. \ref{fig:SI} shows the RMSE performance versus the ratio of the SI power to that of the target echo channel, defined as  $\|\mathbf{H}_\text{SI}\|_F^2/\|\mathbf{h}_\text{t}\mathbf{h}^T_\text{t}\|_F^2$. We observe that variations in SI power have minimal impact on the RMSE performance across a wide range of SI power levels. This robustness is attributed to the assumption of perfect CSI, which enables effective beamforming design to mitigate SI and maintain performance stability. To further investigate the performance in a more realistic scenario, Fig. \ref{fig:ISI} illustrates the RMSE versus the SI channel uncertainty. Specifically, the actual SI channel is modeled as $\mathbf{H}^\text{actual}_\text{SI} = \mathbf{H}_\text{SI} + \widetilde{\mathbf{H}}_\text{SI}$, where $\mathbf{H}_\text{SI}$ is the estimated channel used for optimization, and the estimation error is $\widetilde{\mathbf{H}}_\text{SI}\sim\mathcal{CN}(\mathbf{0}_N,\alpha^2_\text{SI}\epsilon^2\mathbf{I}_N)$ with $\epsilon^2$ representing the level of channel uncertainty. For Fig. \ref{fig:ISI}, the SI/echo power ratio is 50dB. As expected, the RMSE performance of all methods degrades with increasing channel uncertainty. Nevertheless, our proposed methods maintain their superior performance for all values of $\epsilon$. These results demonstrate that while precise SI channel knowledge is essential for optimizing RMSE performance, our methods remain effective in mitigating SI even when channel uncertainties are present.

\section{Conclusion}\label{sec:conclusion}

In this paper, we considered joint array partitioning and transmit beamforming design for monostatic MIMO-ISAC systems. 
We modeled the array partitioning architecture using a binary vector, and then expressed the signals received at the communication users and the sensing receiver. Based on this model, we derived the SINR metric for multiuser communications, and the DOA RMSE metric which involves the 3dB beamwidth and radar SINR.
We then formulated a DOA RMSE minimization problem subject to  constraints on communication SINR, transmit power, and array partitioning. To address the resulting non-convex optimization problem, we developed an alternating ADMM-MM-based algorithm and an efficient heuristic algorithm, each offering different trade-offs between performance and complexity.
Simulation results illustrated that the proposed joint array partitioning and transmit beamforming designs offer notable performance improvements in DOA estimation compared to various benchmark approaches for partitioning the array.
The results highlight the importance of optimally designing the location of the transmit and receive antennas in a partitioned array when designing a monostatic MIMO-ISAC system, and introduce increased opportunities for optimizing the performance trade-off between multiuser communications and target DOA estimation.

Based on this work, future research will focus on addressing practical deployment challenges of the proposed array partitioning and beamforming design in ISAC systems, including RF constraints at higher frequencies, imperfect CSI, computational complexity (especially for extremely large antenna arrays), and trade-offs between performance and hardware consumption.

\end{document}